%% file: main.tex
\documentclass[twocolumn]{aastex}
\newcommand{\about}{$\sim\!\!$~}
\newcommand{\kms}{km s$^{-1}$}

\input{config/theoremstyle}

\begin{document}

\title{S\lowercase{pect}A\lowercase{c}LE: An improved method for modeling light echo spectra}



\author{Roee~Partoush\altaffilmark{1,2}}
\author{Armin~Rest\altaffilmark{1,3}}
\author{Jacob~E.~Jencson\altaffilmark{1,3}}
\author{Dovi~Poznanski\altaffilmark{2,5}}
\author{Ryan~J.~Foley\altaffilmark{4}}
%
\author{Charles~D.~Kilpatrick\altaffilmark{6}}

\author{Jennifer~E.~Andrews\altaffilmark{7}}

\author{Rodrigo~Angulo\altaffilmark{1}}
\author{Carles~Badenes\altaffilmark{8,9}}
\author{Federica~B.~Bianco\altaffilmark{10,11,12,13}}
\author{Alexei~V.~Filippenko\altaffilmark{14}}
\author{Ryan~Ridden-Harper\altaffilmark{15}}
\author{Xiaolong~Li\altaffilmark{8}}
\author{Steve~Margheim\altaffilmark{16}}
\author{Thomas~Matheson\altaffilmark{17}}
\author{Knut~A.~G.~Olsen\altaffilmark{18}}
\author{Matthew~R.~Siebert\altaffilmark{3}}
\author{Nathan~Smith\altaffilmark{19}}
\author{Douglas~L.~Welch\altaffilmark{20}}
\author{A.~Zenteno\altaffilmark{21}}

\altaffiltext{1}{Department of Physics and Astronomy, The Johns Hopkins University, Baltimore, MD 21218, USA} 
\altaffiltext{2}{School of Physics and Astronomy, Tel Aviv University, Tel Aviv, 69978, Israel}
\altaffiltext{3}{Space Telescope Science Institute, Baltimore, MD 21218, USA}
\altaffiltext{4}{Department of Astronomy and Astrophysics, University of California, Santa Cruz, CA 95064, USA}
\altaffiltext{5}{Cahill Center for Astrophysics, California Institute of Technology, Pasadena CA 91125, USA}
\altaffiltext{6}{Center for Interdisciplinary Exploration and Research in Astrophysics (CIERA) and Department of Physics and Astronomy, Northwestern University, Evanston, IL 60208, USA}

\altaffiltext{7}{Gemini Observatory/NSF's NOIRLab, 670 N. A'ohoku Place, Hilo, HI 96720, USA}

\altaffiltext{8}{Department of Physics and Astronomy, University of Pittsburgh}
\altaffiltext{9}{Pittsburgh Particle Physics, Astrophysics, and Cosmology Center (PITT PACC), University of Pittsburgh, Pittsburgh, PA 15260, USA}

\altaffiltext{10}{University of Delaware, Department of Physics and Astronomy, 217 Sharp Lab, Newark, DE 19716, USA}
\altaffiltext{11}{University of Delaware, Joseph R. Biden, Jr. School of Public Policy and Administration, 184 Academy Street, Newark, DE 19716, USA}
\altaffiltext{12}{University of Delaware, Data Science Institute, Newark, DE 19716, USA}
\altaffiltext{13}{Vera C. Rubin Observatory, Tucson, AZ 85719, USA}

\altaffiltext{14}{Department of Astronomy, University of California, Berkeley, CA 94720-3411}

\altaffiltext{15}{School of Physical and Chemical Sciences — Te Kura Mātu, University of Canterbury, Private Bag 4800, Christchurch 8140,
Aotearoa, New Zealand}

\altaffiltext{16}{Vera. C. Rubin Observatory/NSF's NOIRLab, Casilla 603, La Serena, Chile}

\altaffiltext{17}{NSF's National Optical-Infrared Astronomy Research Laboratory, 950 N. Cherry Ave., Tucson, AZ, 85719}

\altaffiltext{18}{NOIRLab, 950 N. Cherry Ave., Tucson, AZ 85719}

\altaffiltext{19}{Steward Observatory, University of Arizona, 933 North Cherry Avenue, Tucson, AZ 85721, USA}

\altaffiltext{20}{Department of Physics and Astronomy, McMaster University, 1280 Main Street West, Hamilton, ON L8S 4M1, Canada}

\altaffiltext{21}{Cerro Tololo Inter-American Observatory/NSF’s NOIRLab, Casilla 603, La Serena, Chile}


\begin{abstract}
Light echoes give us a unique perspective on the nature of supernovae and non-terminal stellar explosions. Spectroscopy of light echoes can reveal details on the kinematics of the ejecta, probe asymmetry, and reveal details on its interaction with circumstellar matter, thus expanding our understanding of these transient events. However, the spectral features arise from a complex interplay between the source photons, the reflecting dust geometry, and the instrumental setup and observing conditions. In this work we present an improved method for modeling these effects in light echo spectra, one that relaxes the simplifying assumption of a light curve weighted sum, and instead estimates the true relative contribution of each phase. We discuss our logic, the gains we obtain over light echo analysis method(s) used in the past, and prospects for further improvements. Lastly, we show how the new method improves our analysis of echoes from Tycho's supernova (SN 1572) as an example.
\end{abstract}

    \input{01Chapters/01Intro}

    \input{01Chapters/02Primers}
    \input{01Chapters/XObservations}
    \input{01Chapters/03Method}
    \input{01Chapters/04Results}
    \input{01Chapters/05Discussion}
    \input{01Chapters/06Conclusion}

    \nocite{*}
    \bibliography{mybib}{}
    \bibliographystyle{aasjournal}

    \appendix
    \input{02Appendices/01CC1}

    \input{02Appendices/02CC2}
    \input{02Appendices/03CC3}

\end{document}

%% file: config/theoremstyle.tex
\newtheorem{definition}{Definition}



%% file: 01Chapters/01Intro.tex
\section{Introduction}
\label{sec:intro}

Scattered light echoes (LEs) of transient events have been studied for over a century \citep[e.g.,][]{1901ApJ....14..167R,Perrine03}, and have been found mainly around recent events. The first person to suggest to use LEs to study ancient supernovae and other transients was \cite{Zwicky40}, however, subsequent surveys to find these LEs were unsuccessful \citep[e.g.][]{vandenBergh65a} due to the fact that they are faint, can be found at large angular separations (several degrees) from the source event, and are often in crowded fields. With the advent of large-format CCDs on telescopes with large apertures enabling deep, wide-field time-domain surveys, systematic searches for faint LEs over large areas are now possible. As part of a microlensing survey \cite{Rest05a} discovered the first LEs of ancient SNe serendipitously, which were associated with several SN remnants (SNRs) hundreds of years old in the LMC. Since LEs in the optical are only marginally attenuated scattered light of the source event, this discovery opened the door for a novel astronomical time machine: we are now able to study these ancient events with modern instrumentation long after their light first reached Earth. 
A study by \cite{Rest08b}, for example, found that the spectrum of an LE associated with SNR~0509-675 is most consistent with particularly high-luminosity Type Ia SN (SN\,Ia). For the first time, an ancient SNR was unambiguously classified using modern spectroscopic capabilities. Additional searches have revealed LEs from Tycho's SN~1572 \citep{Rest08b} and the Cas~A SN \citep{Rest08b,2008Sci...320.1195K}, which were spectroscopically classified as an SN~Ia \citep{2008Sci...320.1195K,Rest11b} and an SN~IIb \citep{Krause2008_TYC}, respectively.

Beyond spectroscopic classification of ancient transients, each unique LE from a single event provides the opportunity to view that object from a different direction \citep{Rest11a,Rest12a}. Using this technique, \cite{Sinnott13} found that H$\alpha$ profiles extracted from SN~1987A's LE spectra show an excess in redshifted emission and a blue knee from the northern hemisphere, while southern hemisphere profiles show an excess of blueshifted H$\alpha$ emission and a red knee. For Cas~A, \cite{Rest11b} found differences in outflow velocities of \about 4000~\kms\ in the He~I $\lambda 5876$ and H$\alpha$ lines from different directions, in agreement with X-ray and optical data of the Cas~A remnant \citep[e.g.,][]{Delaney10,Milisavljevic15}. 
Light echoes thus allow to directly probe the asymmetry of a transient with unprecedented detail.

One of the most basic assumptions used in these studies was that the LE spectrum was the light-curve weighted integrated spectrum of the source event \citep[e.g.,][]{Gouiffes88,Suntzeff88,Schmidt94}. However, this is only valid if the transient time scale is significantly smaller than the time scale associated with light travel through the width of the scattering dust filament. This is often the case to zeroth order for SNe where the transient light curve is short compared with propagation of light through the dust filament, but for events such as Eta Car's Great Eruption, which lasted nearly 2 decades, the opposite approximation is typically valid \citep{Rest12b,Prieto14,Smith18a,Smith18b}. While it is often sufficient for a broad spectral classification to use the assumption of the light-curve weighted integrated spectrum, for more detailed analyses, e.g., subtyping, asymmetry, or spectral time series, it is essential to consider additional factors that contribute to the emission profile in an observed LE spectrum. These include observational factors like slit inclination/position with respect to the LE and astronomical seeing, as well as geometric factors such as the dust filament location, width and orientation.

In this paper, we construct a systematic framework to demonstrate how each of these factors can affect an observed LE spectrum and provide a recipe with which these effects can be characterized. In Section~\ref{sec:method}, we give an overview of the underlying physical setup and concepts, and then describe our algorithm that is built on these assumptions. In Section~\ref{sec:rslt} we apply our method to data in three instructive cases. In Section~\ref{sec:disc} we discuss potential science applications of our method, as well as its limitations. Finally in Section~\ref{sec:conc}, we summarize the improvements in our estimate of the time-varying LE spectrum compared with previous, related works.

%% file: 01Chapters/02Primers.tex

\subsection{Light Echo Geometry Primer}
\label{sec:LE_primer}

\begin{figure}
\epsscale{1.2}
\includegraphics[width=\linewidth]{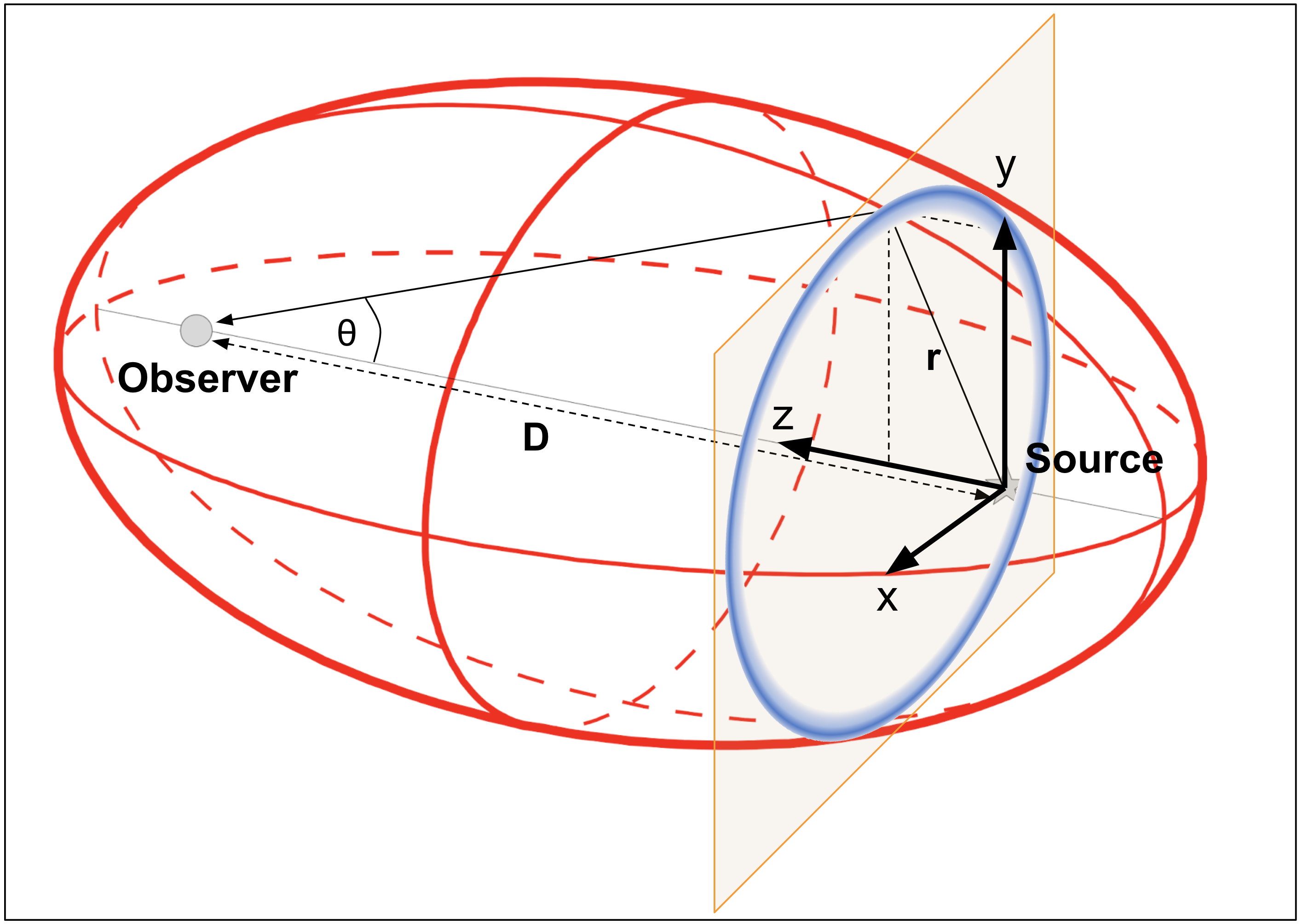}
\figcaption{Geometrical setup of a light echo. Light will be scattered to the observer from any point on the ellipsoid that intersects the dust sheet. For a large enough dust sheet oriented such that its normal points to the observer, this will cause a circular light echo. \label{fig:LE_geometry}}
\end{figure}

LEs can be described and approximated in simple geometric terms (see, e.g., \citealp{2003AJ....126.1939S} for a more detailed review). We define the origin of our coordinate system at the source of the transient event, with the positive $z$-axis toward the observer and the $x$,$y$-axes as the plane of the sky through the event (Figure~\ref{fig:LE_geometry}). Due to symmetry around the line of sight between the event and observer, we can reduce the number of free parameters using $\rho^2=x^2+y^2$, where $\rho$ is the distance of the scattering dust from the line of sight axis. For a given time delay $t$ between the (direct) observation of the light from transient and observation of the corresponding LE, the locus of all possible scattering locations is defined by an ellipsoid whose foci are the source and the observer. Specifically, if $z$ is significantly smaller than the distance $D$ between the event and the observer, then the relation between $z$ and $\rho$ can be approximated with a paraboloid equation known as the famous light echo equation  \citep{1939AnAp....2..271C}:
\begin{equation}
    z = \frac{\rho^2}{2ct} - \frac{ct}{2}\label{eq:LE_eq}
\end{equation}
We note that the coordinate $\rho$ can also be calculated  from the angular separation $\theta$ as $\rho=(D-z) tan{\theta}$, if $D$ is known.

\begin{figure}
\epsscale{1.2}
\includegraphics[width=\linewidth]{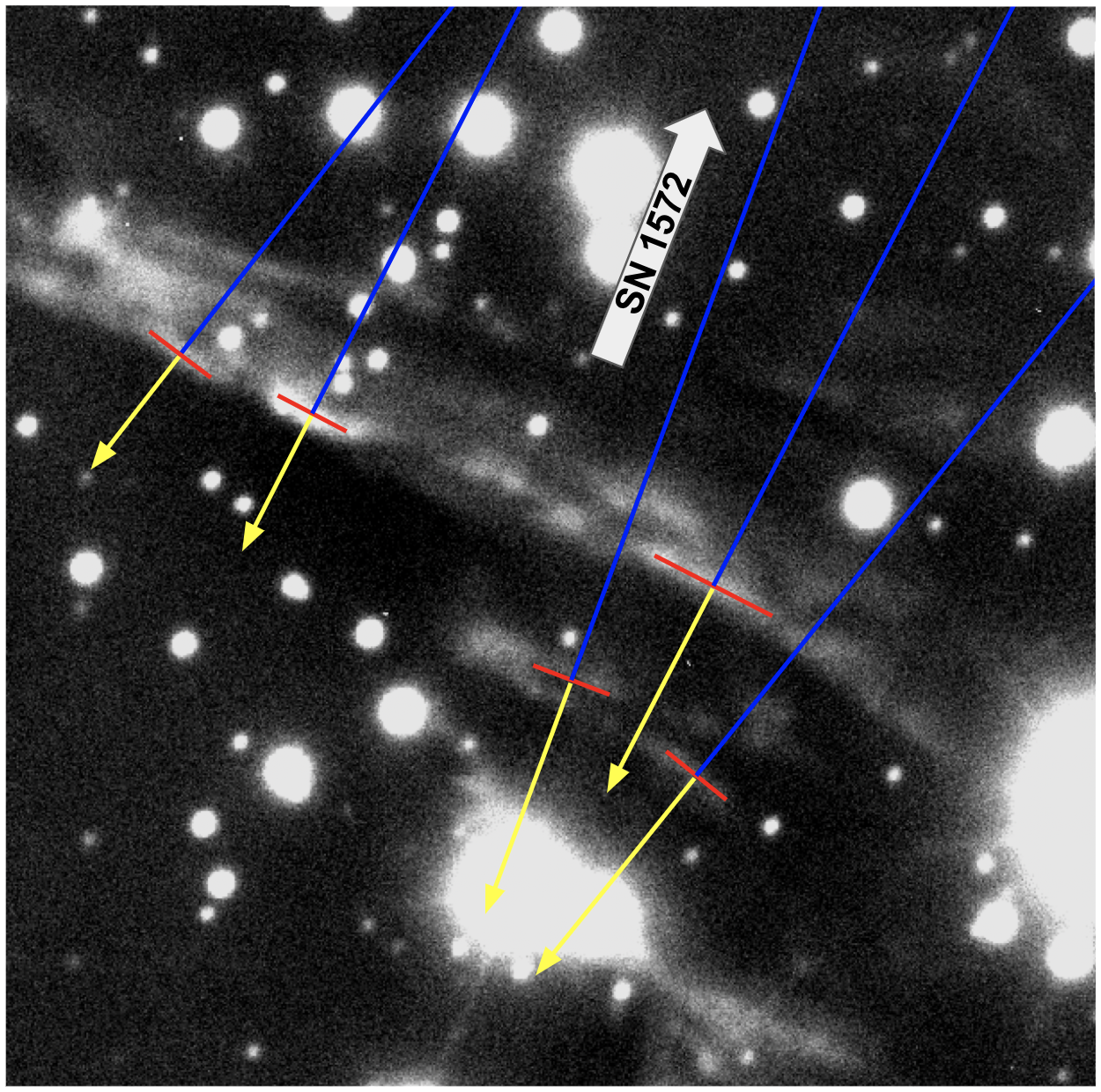}
\figcaption{The LE crest is the line of maximum brightness in the LE (red lines). The LE motion vector (the direction perpendicular to the crest yellow arrows) usually points roughly in the direction opposite to the source event (blue lines). The true direction to the source event, SN\,1572, is indicated at the top by the large white arrow.
\label{fig:LE_cMv}}
\end{figure}

The light echo equation explains some of the observed characteristics of light echoes. We see a light echo at time $t$ from any dust that intersects the associated paraboloid as defined in Equation~\ref{eq:LE_eq}. If the scattering dust is a dust sheet perpendicular to the line of sight, the intersection is a circle. This is the reason that light echoes often resemble circles \citep[for example the famous light echoes of SN~1987A][]{Rest05b} or arcs if the dust dimensions are smaller than $2\rho$.  Since $\rho$ increases with increasing $t$, the circles or arclets appear to move away from the event source with time. This means that the LE motion vector (defined as the vector normal to the LE crest pointing in the direction of the light echo motion, as shown in Fig.~\ref{fig:LE_cMv}) extrapolated backwards goes through the source event position. In other words, the position angle (PA) of the LE motion vector is the same as the PA between the light echo and the source event.

This is of course an idealized scenario. For example, if the scattering dust sheet or filament is rotated around the $\rho$-axis, the crest of the light echo arclet also rotates. This, in turn, means that the PA of the LE motion vector is no longer the same as that between the LE and source event. 
Fig~\ref{fig:LE_cMv} shows an example of a LE group in which individual LEs have  different LE motion vectors. However, we note that these differences are in general small ($\lesssim$few deg). 
This has been used in the past to associate LEs with their source events by extrapolating the motion vectors backwards \citep[e.g.,][for Cas A and Tycho]{Rest08b}.
If the dust filament has a rotation toward or away from the observer, the LE apparent motion increases and decreases, respectively. Thus the geometry of the scattering dust filaments 
may have a significant impact on the observed LE properties.
In Section~\ref{sec:LESpc_primer} below, we discuss how these effects then impact spectroscopy of LEs.
Table~\ref{tab:core_def} summarizes several key definitions that will be used throughout the paper.


\begin{table*}[ht]
  \centering
  \resizebox{\textwidth}{!}{
  	\begin{tabular}{ |l|l| } 
		\hline
		\textbf{Name} & \textbf{Description} \\
		
		\hline
            LE crest & A line that intersects the points on the observed LE that scatter light from phase zero (peak brightness), see Fig.~\ref{fig:LE_cMv}. \\
            Projection scale ($P$) & The ratio between angular displacement and the difference in corresponding phase along the source event's light curve.\\
            LE motion vector & The direction perpendicular to the LE crest.\\
            LE profile & brightness vs. angular displacement along a given direction (defined by the LE profile box).\\
            LE profile model & a light curve convolved with a Gaussian kernel representing dust and seeing effects.\\
            LE spectrum & spectroscopic measurement of an observed LE.\\
            LE spectrum model & a weighted sum of real measured spectra from SNe at different phases.\\
            Effective light curve & The weights used for synthesizing the LE spectrum model from a spectrophotometric library.\\
	    \hline
	\end{tabular}
    }
  \caption{Key definitions}
  \label{tab:core_def}
\end{table*}


\subsection{Light Echo Spectroscopy Primer}
\label{sec:LESpc_primer}

Sufficiently bright LEs can be used to obtain spectroscopic observations of the source event for which the original, unscattered light first reached Earth long before modern instrumentation was available. 
In contrast to spectroscopy of current transients, the spectrum of an LE will contain flux from more than one epoch. The reason is that each phase of the source transient is 
reflected simultaneously, with the LE flux of a given phase proportional to the intrinsic brightness of the transient at that phase.
This means that if all LE flux is captured in a spectrum, the resulting LE spectrum is the light-curve-weighted integrated spectra at all phases. This assumption has been used in the past to classify and interpret LE spectra, e.g., for the classification of SNR\,0509-67.5 in the LMC \citep{Rest08a}. 

In practice, an observed spectrum will only capture a portion of the LE flux, simply because the slits used in spectroscopic observations have finite angular sizes. As the arrival time for the direct emission from a transient, $t_0$, is different for each phase, the corresponding light echo ellipsoids vary in size. This means that the LE for each phase is mapped to a slightly different position on the sky. Some phases, therefore, are not captured by the slit and will not contribute to the observed LE spectrum \citep{Rest11a,Rest11b,Rest12a}. This is further complicated by the smearing of the LE from the finite width of the dust sheet and the optics/atmosphere during observations. The end result is that observed LE spectrum is the time-integrated spectrum of the transient weighted by an effective light curve, defined as the original light curve modified by a window function that accounts for the geometry of the scattering dust and observational parameters such as the slit orientation, width, and the point spread function (PSF). 

In many cases, the effective light curve is similar to the original light curve, but with early and late phases cut off \citep[e.g.,][]{Rest11b}. As a result, the assumption that LE spectra are the light-curve-weighted integrated spectra of the source event often works reasonably well for an overall spectral classification. However, for more detailed LE analyses that rely on on subtle differences in spectral features, such as 3D spectroscopy or sub-classification, it is essential to know the effective light curve accurately. 
The challenge is that it is difficult to independently determine the geometric properties of the scattering dust and observational parameters, which are often degenerate. Here, we have developed a forward modeling method that empirically determines and accounts for the two critical parameters for a given LE observation: the mapping of the phases of the source event to spatial coordinates on the sky and the combined smearing effects of seeing conditions and dust properties. 

\begin{figure}
\epsscale{1.2}
\includegraphics[width=\linewidth]{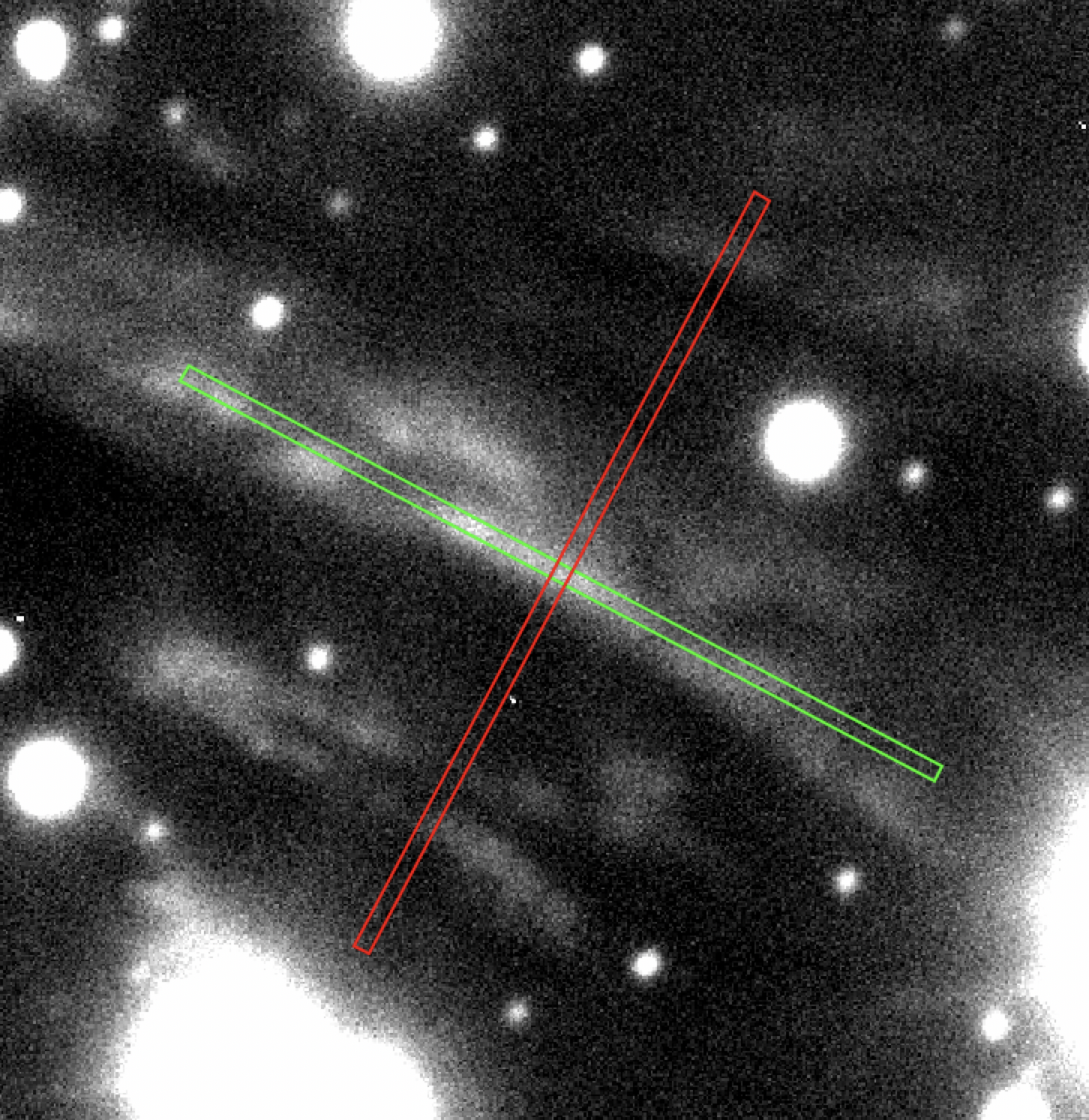}
\figcaption{LE spectroscopy - slit orientation has a significant effect on the measured LE spectrum. Here we see two example orientations - a red slit perpendicular to the LE crest, and a green slit parallel to the crest.
\label{fig:LESpc_primer}}
\end{figure}


\begin{figure*}
\epsscale{1.2}
\includegraphics[width=\textwidth]{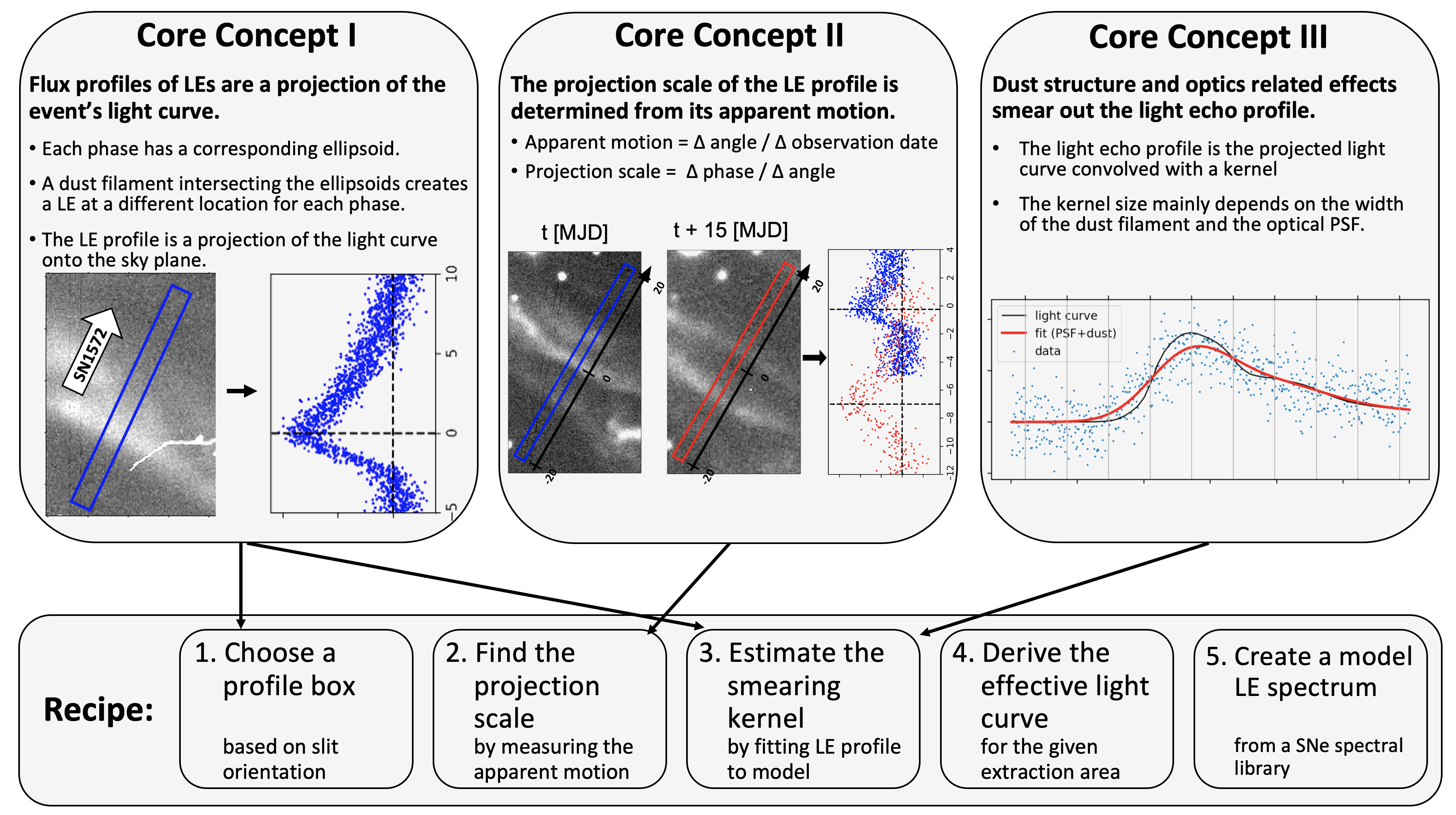}
\figcaption{LE spectra modeling---a bird's eye view. The three boxes on the top lay out the core concepts, with corresponding section numbers. The bottom box gives an overview of the recipe, as described in Section~\ref{sec:recipe}.
\label{fig:one_big_figure}}
\end{figure*}

%% file: 01Chapters/XObservations.tex
\section{Observations \& Data Reduction}
\label{sec:observation}

We use images from various observing runs on the Mayall 4m telescope at Kitt Peak National Observatory from 2009 to 2015. We used the Mosaic-1 and Mosaic3 imagers, which operate at the f/3.1 prime focus at an effective focal ratio of f/2.9. For maximum depth, we use the Bernstein VR Broad filter (k1040) which has a central wavelength of 594.5nm and a FWHM of 212.0nm. We also use images from observing runs at the Keck observatory using LRIS \citep{oke95} and DEIMOS \citep{Faber_KeckDeimos03} on 2012 and 2013.
Both the KPNO and Keck images are reduced using the photpipe pipeline \citep{Rest05a,Rest14}, which is a robust and well-tested difference image pipeline. Photpipe applies standard reduction (bias subtraction, flat-fielding), aligns the WCS, and performs photometry using Dophot \citep{Schechter93}. Then the absolute photometric calibration is done using the PanSTARRS1 catalog \citep{Chambers16}. The difference image is created using Hotpants \citep{2015ascl.soft04004B}\footnote{https://github.com/acbecker/hotpants}, which determines the appropriate spatially varying convolution kernel to match a template.

%% file: 01Chapters/03Method.tex
\section{Method}
\label{sec:method}

\begin{figure*}
\epsscale{1.2}
\includegraphics[width=\textwidth]{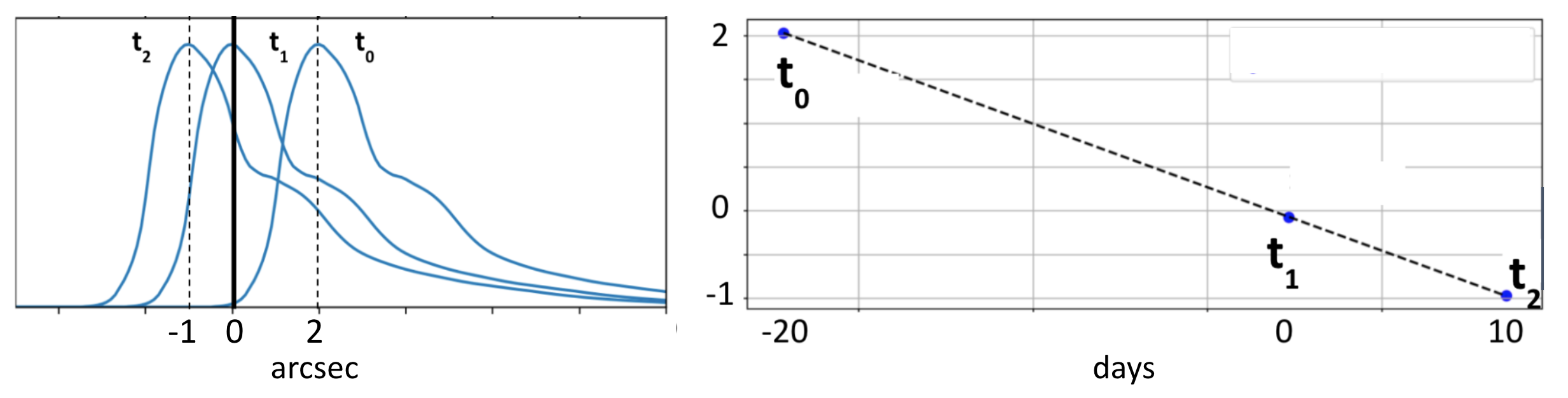}
\figcaption{Apparent motion and projection scale. \textit{Left:} LE profiles for three consecutive observations of the same LE on different dates, plotted on a single graph. \textit{Right:} peak location vs. observation date. The slope of the line of best fit is the apparent motion, which is inverse to the projection scale.
\label{fig:appmotion_illu}}
\end{figure*}

\begin{figure*}
\epsscale{1.2}
\includegraphics[width=\textwidth]{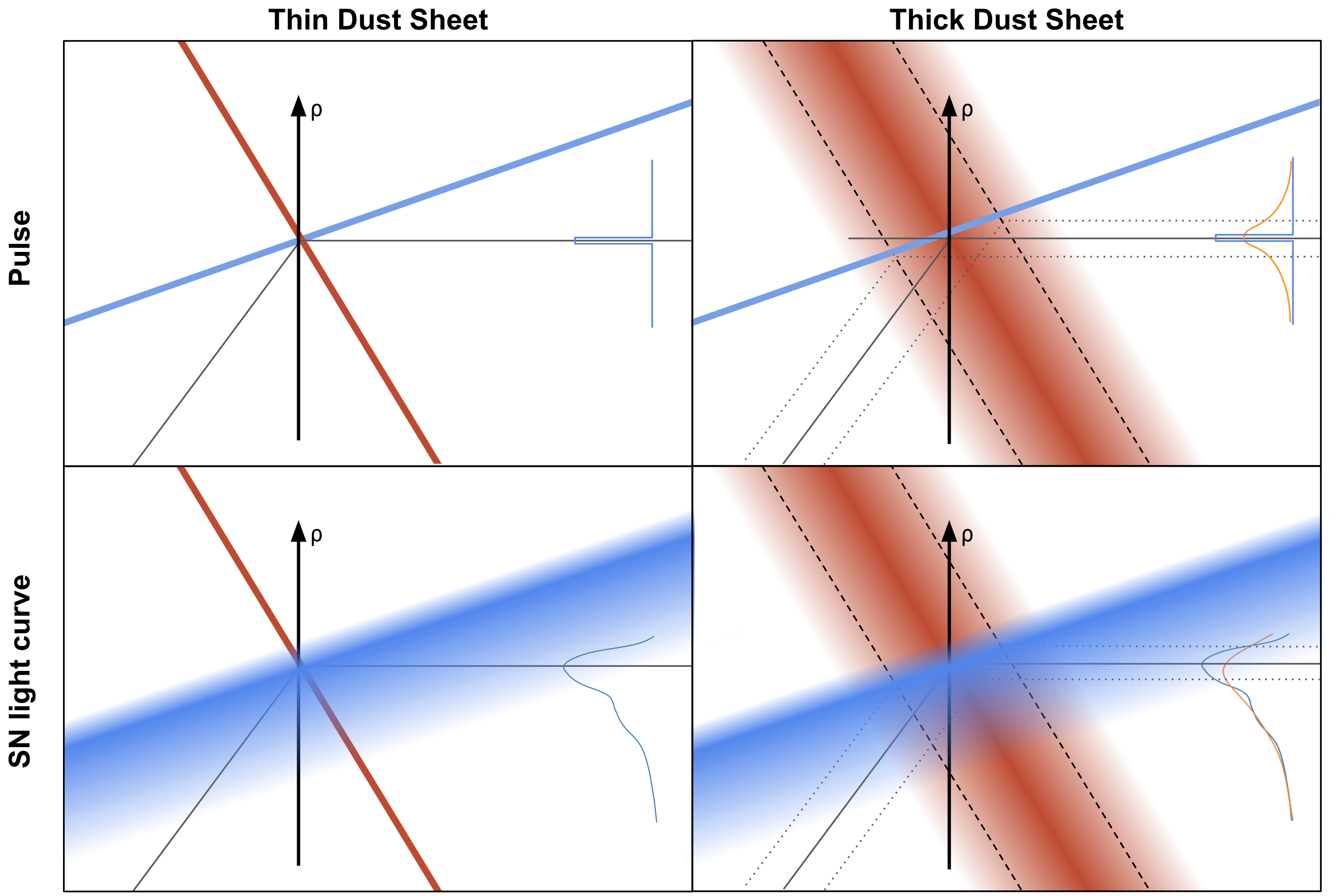}
\figcaption{Dust-related phase spread, neglecting PSF-related spread. The top two panels assume a delta function light curve. \textit{Top Left:} a delta function light curve and a thin dust sheet (delta function density distribution) produces a delta function flux profile (left panel). \textit{Top Right:} a thick dust sheet (Gaussian density distribution) with a delta light curve produces a flux profile identical in shape to the density profile of the dust sheet, $G(\sigma_{\rm dust})$. The bottom two panels show an SN Ia light curve (blue shaded region). \textit{Bottom Left:} for a very thin dust sheet, the flux profile is simply a projection of the light curve. \textit{Bottom Right:} a wider dust sheet would cause a phase spreading effect (orange curve), as explained in Section~\ref{sec:mthd_CC3}. \label{fig:dust_width}}
\end{figure*}

The basic procedure in LE spectral analysis is to compare template spectra derived from observations of well-observed extragalactic SNe to the observed LE spectrum. In previous studies, the primary assumption was that a LE spectrum is the light-curve weighted integrated spectrum of the source event. However, this is only an approximation, and it is not sufficient if we want to derive physical properties of the underlying transient beyond simple spectral classification. Many factors have an impact on which phases of the source event contribute to the observed light echo spectrum, but they can all be summarized and consolidated into an effective light curve (as opposed to the full light curve) that is used as the weight when constructing the spectral templates.

In this Section, we describe the recipe to construct this effective light curve. We first describe three core concepts which are essential to understanding the principles of our forward-modeling method. An overview of the core concepts and how they apply to the recipe is shown in Figure~\ref{fig:one_big_figure}.
Code implementation of the method and example cases can be found at the GitHub repository \texttt{SpectAcLE} \footnote{\url{https://github.com/RoeePartoush/SpectAcLE}}.

\subsection{Core Concept 1: Flux profiles of light echoes are projections of the event's light curve.}
\label{sec:mthd_CC1}
Consider this very simple spatial configuration of the scattering dust: a perfectly planar and infinitesimally thin dust sheet perpendicular to the line of sight, extending far enough to intersect a full circle with an ellipsoid corresponding to a certain time delay $t_{0}$ between the date when light from that phase initially reached Earth and the LE observation date (as described in Section~\ref{sec:LE_primer}). For an observation at a given time $t_{obs}$, the light echo we see is then a circle, the cross section of the dust sheet with the light echo ellipsoid. Considering that the transient source event has some finite duration, each phase has a slightly different delay time, all varying around $t_{0}$. Thus, for a given $t_{obs}$, each phase has a different light echo ellipsoid, with earlier phases having larger ellipsoids and later phases having smaller ellipsoids. 
Thus, each phase produces a distinct light echo circle of a specific radius, with earlier phases associated with larger circles (greater separation from the source). The brightness of each circle component of the LE is proportional to the intrinsic brightness of the source event at that phase. We refer to the light echo profile as the one-dimensional flux profile slicing through the light echo in the direction of the source event, 
and define the projection scale as the conversion ratio from the phase of the source event to the spatial coordinate along the LE profile on which light from that phase is scattered.
For this idealized example, the light echo profile is simply the projection of the event's light curve. The upper left panel of Figure~\ref{fig:one_big_figure} contains an actual example of a light echo and its profile. For a more detailed derivation and explanation of this concept see Appendix \ref{sec:XCC1}.

\subsection{Core Concept 2: The projection scale is the inverse of the apparent motion}
\label{sec:mthd_CC2}

The 
projection scale is an essential component of interpreting LE spectra. In the idealized case described in Core Concept 1, the projection scale can be derived geometrically. However, in practice, the scattering dust sheet or filament is rarely perpendicular to the line of sight. Rotation of the dust sheet toward and away from the observer (rotation around x-axis in Fig.~\ref{fig:LE_geometry})  ``stretches'' and ``squeezes'' the light echo profile, respectively. Rotation around the $\rho$-axis rotates the principle light echo axis with respect to the direction toward the source event, adding further complications. In addition, the PA of the light echo profile extraction box is often set by outside parameters, e.g., the PA of the slit used for obtaining a spectrum of the light echo, which means that this box is not perpendicular to the principle axis of the light echo. Since we rarely know the 3D properties of the scattering dust to sufficient accuracy a priori, we cannot geometrically derive the projection scale. Fortunately, it can obtained empirically to high accuracy: as described in more detail in Appendix~\ref{sec:XCC2}, the projection scale is simply the inverse of the observed apparent motion of the LE flux profile (see Fig.~\ref{fig:appmotion_illu}). 
In geometric terms, the light echo ellipsoid for the event peak at observation time $t_{obs}$ is exactly the same as the light echo ellipsoid at phase $\Delta t$ observed at time $t_{obs}-\Delta t$.

\subsection{Core Concept 3: Dust and seeing induce phase-smearing in the light echo profile}
\label{sec:mthd_CC3}

We must also consider the effects of a dust filament of finite width and observational effects, including astronomical seeing and telescope/instrument optics. When the dust is no longer thin, a given LE ellipsoid will intersect the dust filament in multiple locations, which are in turn projected onto a range of observed positions on the sky. This leads to a smearing of the light echo profile (as illustrated in Figure~\ref{fig:dust_width}, which can be adequately modeled as a convolution with a Gaussian of width, $\sigma_{\mathrm{dust}}$. Similarly, the effects of seeing and finite optics can be captured with an effective point-spread-function (PSF), modeled as as second convolution with a Gaussian of width $\sigma_{\mathrm{PSF}}$. As described in more detail in Appendix~\ref{sec:XCC3}, these effects can therefore be combined into a single convolution with a Gaussian of total width $\sigma^2_{\mathrm{tot}} = \sigma^2_{\mathrm{dust}} + \sigma^2_{\mathrm{PSF}}$. 

With these three core concepts in mind, we use forward modeling to determine the effective light curve for a given slit extraction. We first define a LE profile box that is appropriate for the given slit and determine the projection scale from the apparent motion of the LE along the box. Using the projection scale to transform angle into time, we model the effects of the dust and PSF on the light curve as convolution with a Gaussian, and fit this model to the observed light echo profile to estimate the dust width. The effective light curve is then derived using the exact position, orientation and width of the spectrograph slit. In the next section, we describe this process in detail.

\subsection{Recipe}
\label{sec:recipe}
The process of modeling a LE spectrum can be broken down into four steps: (1) choosing a profile box, (2) estimating the projection scale, (3) estimating the dust width, and finally (4) calculating the effective light curve.

\subsubsection{Step 1: Choosing a LE profile box.}
When choosing a profile box, we distinguish between two cases: (1) the slit is roughly perpendicular to the crest of the LE ($\alpha \le 45^{\circ}$), and (2) the slit is roughly parallel to the crest of the LE ($\alpha > 45^{\circ}$).
In the first case, the light echo profile box is chosen to have the same angle as the slit (see Figure~\ref{fig:rslt1_1}). In the second case, a set of several parallel LE profile boxes that are perpendicular to the slit is more appropriate (e.g., the case described in Section~\ref{sec:rslt_3}).
Next, we choose the width of the box profile. Here, we have two competing considerations: the precision and the accuracy of the model. On the one hand, a wider box leads to a higher signal-to-noise ratio, and we can obtain a more precise estimation of the parameters. On the other hand if the box is too wide, the model we derive may not accurately represent the specific part of the dust sheet that was probed by the slit but instead an average over the wider box profile.

\subsubsection{Step 2: Finding the projection scale.}
To find the projection scale $P$ of the LE profile, we measure the apparent motion $\mu$ of the LE along the profile box in a time-series of images.
This is done by finding the positions of peak brightness in the sequence of LE profiles,
and then finding the line of best fit using linear regression.
The slope of this line is the apparent motion $\mu$ of the light echo, which is the inverse of the projection scale $P$ (see right panel in Figure~\ref{fig:appmotion_illu}).

This process requires at least two consecutive images (and preferably three, to validate the assumption of constant apparent motion) of the same area observed over a period spanning the time before and after the spectroscopic observation, preferably with time intervals roughly equal to the duration of the source event (e.g., several months to a year for a typical SN\,Ia). Achieving a good linear fit under these requirements will validate that the apparent motion is constant over the span of the extraction area on the slit.

\subsubsection{Step 3: Estimating the smearing kernel.}
\label{sec:recipe_step3}
To calculate the smearing Gaussian kernel described above in \ref{sec:mthd_CC3}, 
we fit the observed LE profile to a model LE profile, which is essentially the result of a convolution between two signals: (1) a template light curve of an observed event (e.g., a typical SN\,Ia from the SALT library \citealp{Guy2005}), and (2) the smearing kernel, denoted by $K_{t_s}$, at the time the LE spectrum was observed (\ref{fig:rslt1_2}, top panel). Note that the smearing kernel models the combined effect from both a finite dust width and the optical PSF. 
We emphasize that the projection scale calculated in the previous step has enabled us to align the light curve (brightness over time) to the LE profile (brightness over space). 

At this point it is important to note one consideration: which image to use for estimating the kernel. It is preferable to use an image that was observed right before or after the spectroscopic measurement, and thus had approximately the same seeing conditions. However, a high S/N is also a priority in this step. Unlike in Step 2, where we only needed to obtain the location of the peaks in the LE profiles to fit the apparent motion, here the specific shape of the profile affects the result. If a high S/N image from the same night as the spectroscopy observation is not available, the smearing kernel has to be broken into its two components: the dust width and the PSF. Since the PSF is known, the fitting process allows us to obtain the dust kernel. The dust kernel, together with the PSF from the spectroscopic observation, can then be used to compute the effective light curve.

\subsubsection{Step 4: Calculating the effective light curve.}
Using the projection scale and the smearing kernel from the previous steps, along with the extent of the extraction area, or slitlet, along the slit that was used to obtain the LE spectrum, we can produce the effective light curve. The effective light curve will be used to sum together template SN spectra from different phases with the correct weights, to produce an accurate model of the observed LE spectrum.
We denote the length of the slitlet as $b_1$. Using the projection scale, we calculate the corresponding temporal extent of the slitlet, i.e., the range of phases probed by the slitlet. 
This is modeled as a box-car function denoted $B(b_1)$. Next, we calculate the convolution of the slitlet aperture with the smearing kernel $K_{t_s}$ obtained in the earlier step, to produce the weight function $W_{t_s}$.
\begin{eqnarray}
W_{t_s} & = & K_{t_s}  \ast B(b1)  \label{eq:weight_fun}
\end{eqnarray}

\begin{figure*}
\epsscale{1.0}
\includegraphics[width=\textwidth]{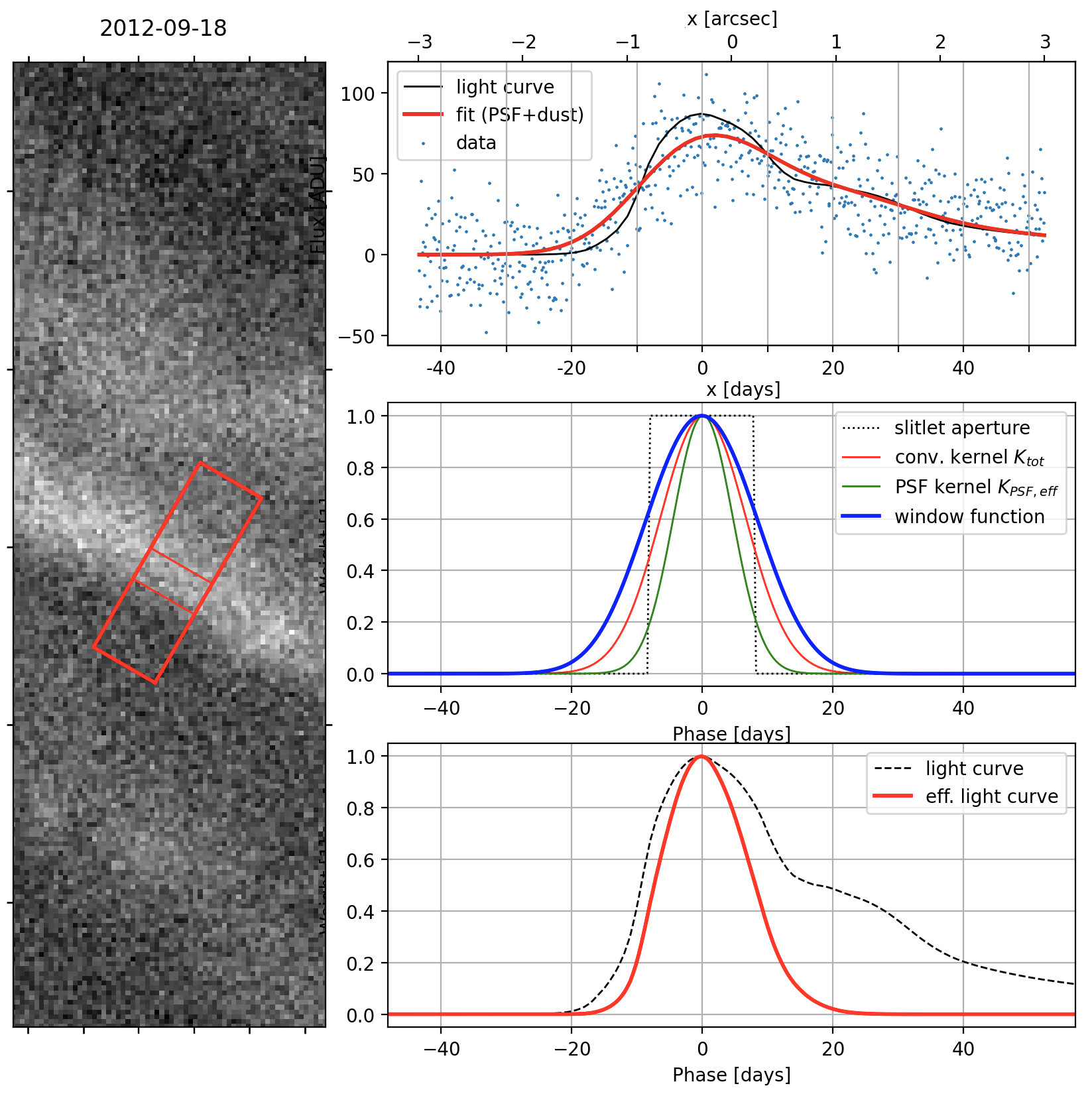}
\figcaption{An example case showing steps 3 and 4 of the analysis process. \textbf{Left: } the image used for profile fitting. The red overlay indicates a 6$\arcsec$~long and 2$\arcsec$~wide profile box used in the fitting the LE profile, and also a 1$\arcsec$ slitlet aperture centered on the peak. To the right, the upper panel shows the LE profile model of best fit along with the measured LE profile data. The middle panel illustrates the derivation of the weight function, and the bottom panel shows the calculation of the effective light curve for the given slitlet and LE of interest. \label{fig:rslt1_2}}
\end{figure*}

The weight function $W_{t_s}$ represents the relative contribution, i.e., the weight, of each phase of the SN in the LE spectrum as estimated by our modeling process.
The effective light curve $L_{eff,t_s}$ is then obtained simply by multiplying the weight function by the template light curve (Figure~\ref{fig:rslt1_2}, bottom panel).
Using the effective light curve, we can sum template spectra from different phases of the SN to produce the LE spectrum model.

%% file: 01Chapters/04Results.tex
\section{Results}
\label{sec:rslt}

\begin{figure*}
\epsscale{1.0}
\includegraphics[width=\textwidth]{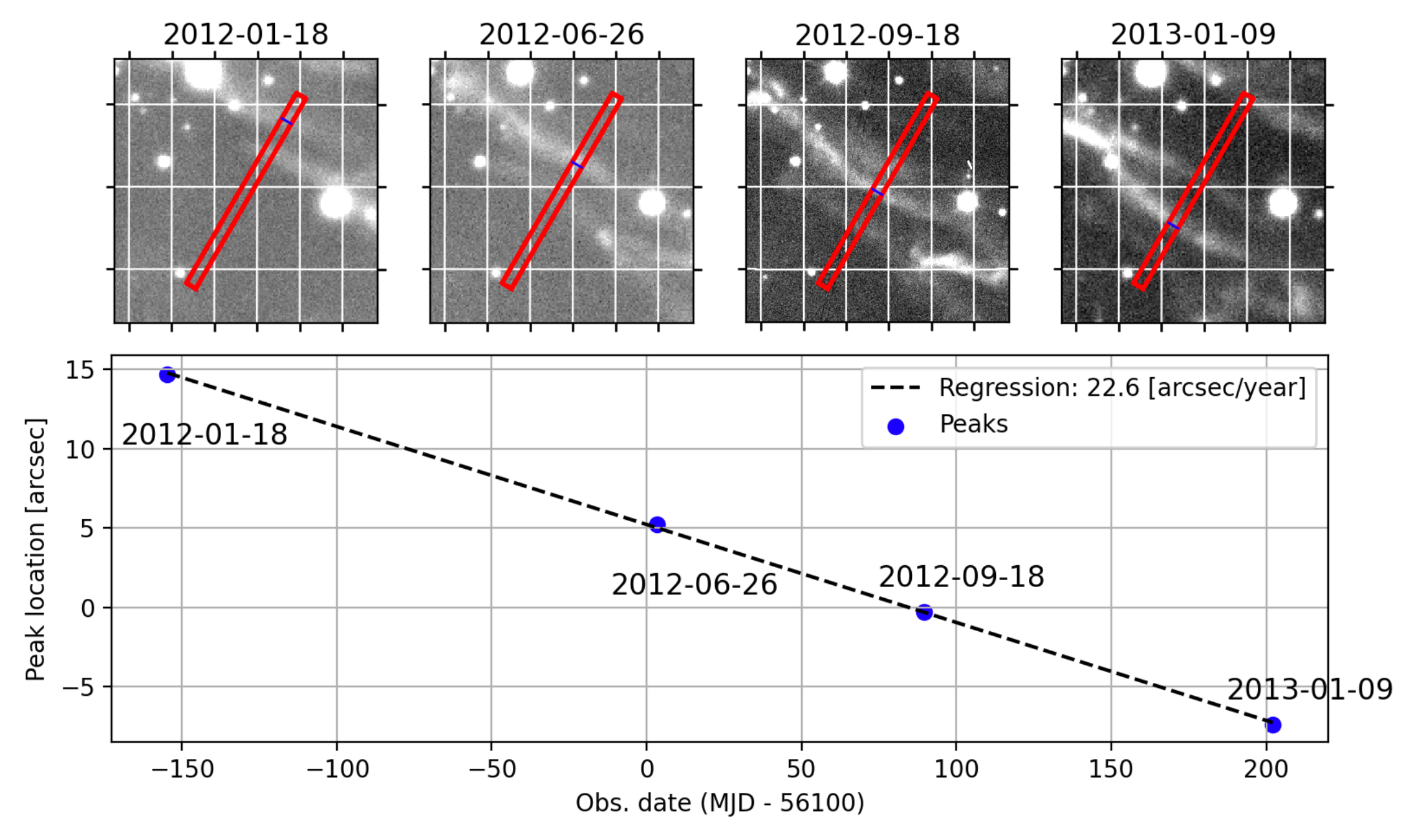}
\figcaption{Case 1: Linear fit for apparent motion $A$. The top panels show 4 images of a light echo spanning 1 year taken with Kitt Peak Mayall Mosaic 3 and Keck Low-Resolution Imaging Spectrograph (LRIS). The red box indicates the sky region from which the light echo profile is extracted. The blue marker indicates the peak of the light echo profile obtained by curve fitting. The lower panel shows these fitted peak locations versus the observation date. We estimate the uncertainties of the peak locations to be \about0.2 arcsec. The fit is excellent, and the slope $A=22.5$~arcsec~yr$^{-1}$ is the apparent motion of this light echo. We can then directly derive the projection scale as $P=16.2$~days~arcsec$^{-1}$ using Equation~\ref{eq:P_V}.
    \label{fig:rslt1_1}}
\end{figure*}

We present here the results for three cases where we have employed the method described above to produce an effective light curve for a given slit using a time series of images containing light echoes traversing across the sky.
All LEs presented in this Section originate from Tycho's SN (SN\,1572), imaged on the same field on the sky using the Mayall 4-m telescope at Kitt Peak observatory, and the Low-Resolution Imaging Spectrograph (LRIS) and Deep Imaging Multi-Object Spectrograph (DEIMOS) instruments at Keck observatory.

The three cases present a variety of uses and circumstances. The first case (Section \ref{sec:rslt_1} is fairly simple --- a single, planar dust structure, a slit orientation very nearly perpendicular to LE crest, and small PSF. Such a case does not require any deviations from the process prescribed in \ref{sec:recipe}.
The second case (Section \ref{sec:rslt_2}, on the other hand, is an example where our method does not work ``by the book'' due to a complex dust structure and thus requires certain modifications to the general method.
The third case (Section \ref{sec:rslt_3} is more complicated still --- the extent and orientation of the slit are such that we chose to divide the slit into multiple slitlets and model the flux profile for each of them.





\subsection{Case 1: Vanilla Light Echo}
\label{sec:rslt_1}



In this case our data is a series of four images taken over the course of one year, shown in Figure~\ref{fig:rslt1_1}. We choose a profile box oriented roughly perpendicular to the LE crest. As described in \ref{sec:recipe}, our first step is calculating the projection scale for the direction defined by the profile box.  The locations of those peaks, combined with the observation dates, allow us to calculate the apparent motion of the LE to be 22.6 arcsec~yr$^{-1}$ and the corresponding projection scale of the LE profiles to be 16.1~days~arcsec$^{-1}$.

Once we have the projection scale, we can fit the model to the flux profile extracted from each image (this time with the projection scale held fixed at 16.1 days~arcsec$^{-1}$). The value of $\sigma_{\rm dust}$ resulting from the fit is an estimate for the effective dust related phase spread. In principle, any of the four images in the series can be used to obtain $\sigma_{\rm dust}$. We choose the third image in the series (Keck/LRIS, 2012 September 18) as this image has the best seeing and depth among the four.

We find a dust width of \(\sigma_{\rm dust}\)=12 days, which along with a measured FWHM of 0.67~arcsec (measured from stars present in the same image) and the estimated projection scale, gives a total phase-smearing kernel of 16~days. We can now calculate the window function to be used in forward modeling of a given spectroscopic measurement. For example, we can take a 1~arcsec long extraction area centered on the peak brightness of the LE. The window function is given as the convolution of the phase-smearing kernel --- a Gaussian of total width $\sigma_{\rm tot} = \sqrt{\sigma_{\mathrm{PSF}}^2 + \sigma_{\rm dust}^2}$ (see \ref{sec:mthd_CC3}) --- with a boxcar function corresponding to the slitlet aperture (Equation~\ref{eq:weight_fun}). In this case (assuming same PSF for the spectroscopic measurement and the image used for fitting) we have a kernel with FWHM of 16~days and a boxcar function 1~arcsec wide.  Then, the effective light curve is given by multiplying the template light curve by the window function (see Figure~\ref{fig:rslt1_2} given earlier as an example, middle and bottom panel on the right). The effective light curve gives the relative weights used to properly sum template spectra from a range of phases to produce the forward-modeled spectrum for the given slitlet.


\subsection{Case 2: Complex Dust Structure}
\label{sec:rslt_2}

\begin{figure*}
\epsscale{1.0}
\includegraphics[width=\textwidth]{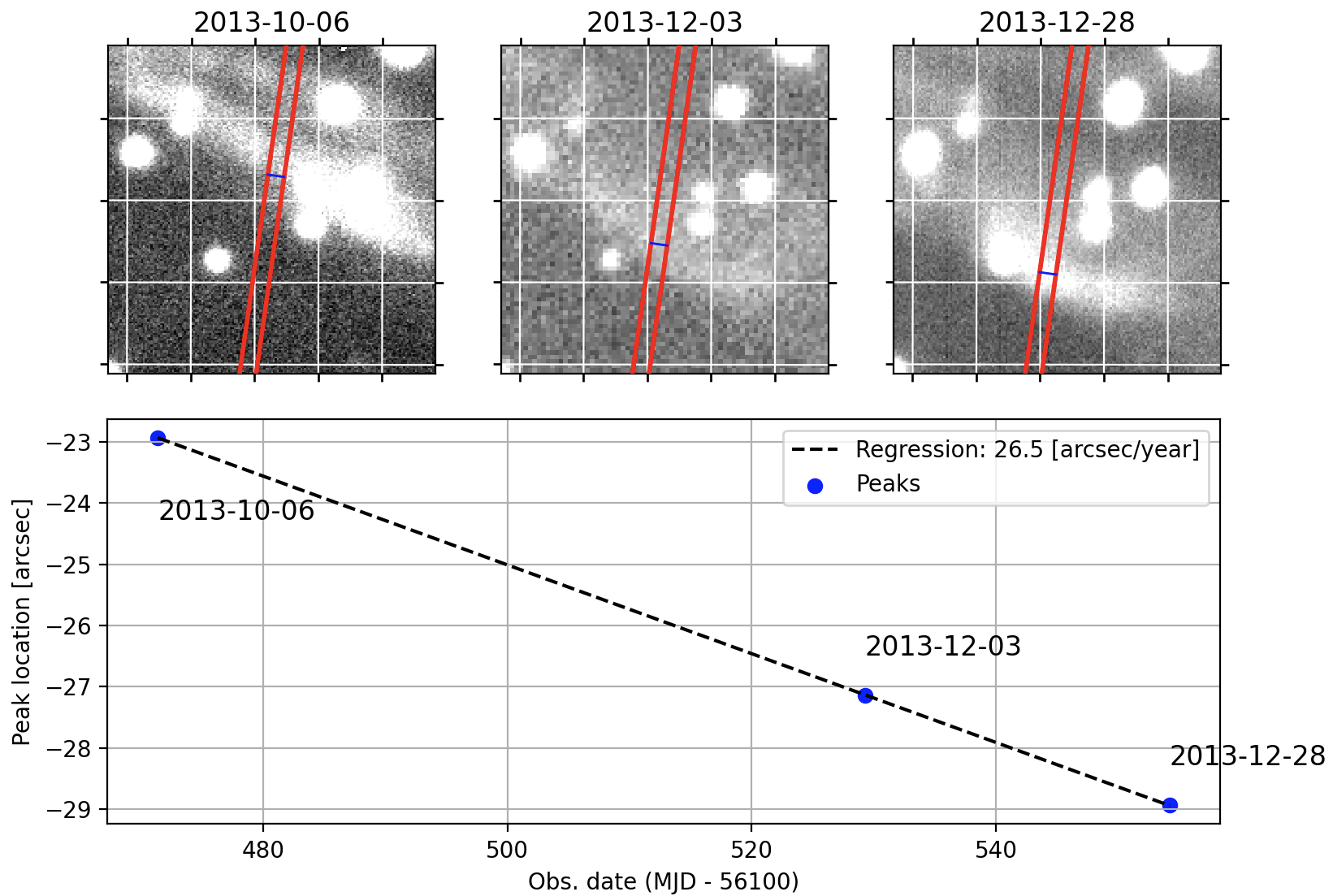}
\figcaption{Case 2: apparent motion estimation by linear fit, images and a plot showing peak location vs observation date. Markings same as in Fig.~\ref{fig:rslt1_1}
\label{fig:rslt2_1}}
\end{figure*}

\begin{figure*}
\epsscale{0.9}
\includegraphics[width=\textwidth]{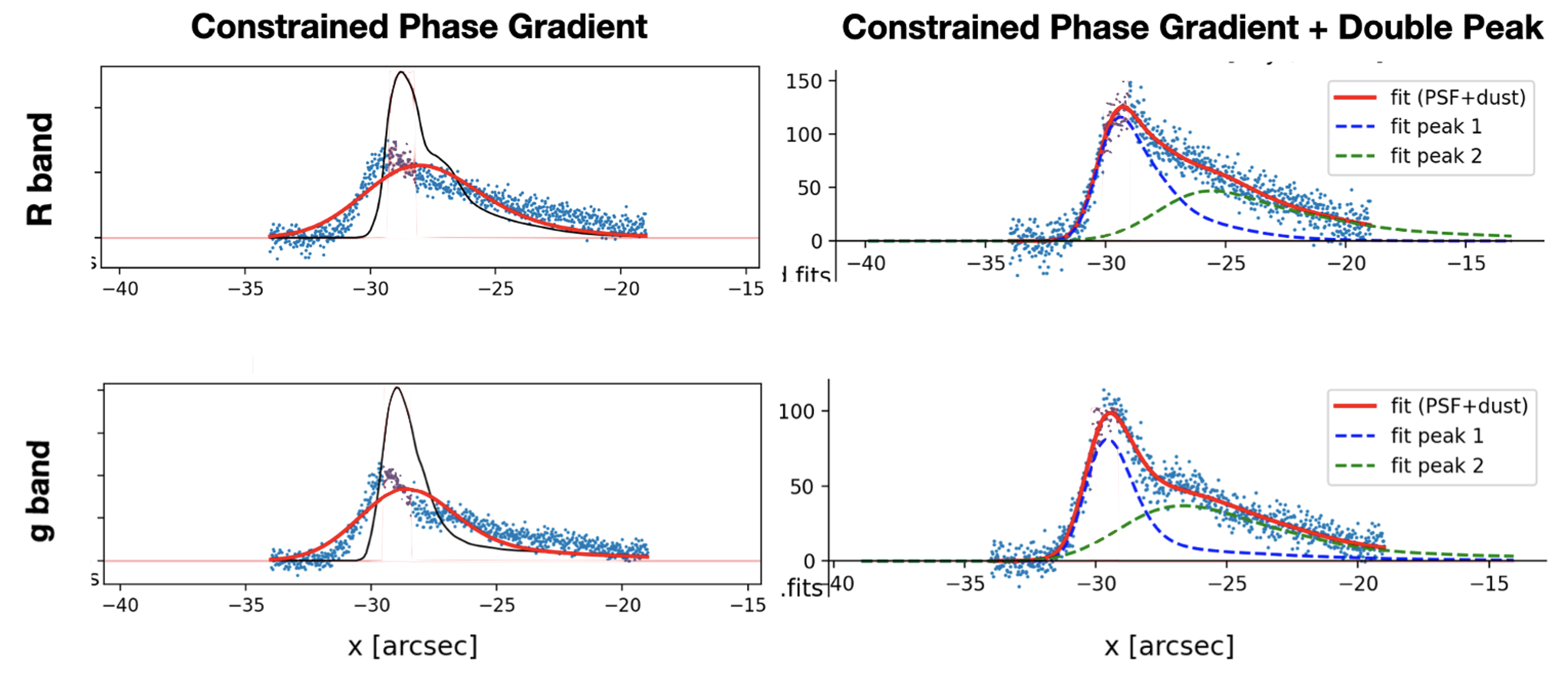}
\figcaption{Case 2: Fitting flux profiles from Keck LRIS 2013-12-28, $R$ and $g$ filters, with three different LE profile models. 
\label{fig:rslt2_2}}
\end{figure*}



In this case we have a time series consisting of three images spanning a three months period, as shown in Figure~\ref{fig:rslt2_1}, and the spectroscopic measurement to be modeled was measured on the same night the third image was taken. As in the previous case, by extracting the peak locations we find an apparent motion of 26.5~arcsec~yr$^{-1}$, or equivalently a projection scale of 13.8~days~arcsec$^{-1}$.

However, fitting the LE profiles to our LE profile model with the projection scale set to 13.8~days~arcsec$^{-1}$ 
does not accurately capture the flux profile of the LE, as can be seen in Figure~\ref{fig:rslt2_2}, left column.
Even when fitting the LE profiles with the projection scale as a free parameter, we are not able to accurately reproduce the measured LE profile.


To resolve this, we try a modified LE profile model, which is a sum of two single-sheet flux profile models. This model describes the behavior we would expect in case we had two separate dust sheets in parallel to each other: the ellipsoidal intersections of the two sheets as seen from Earth can overlap, in which case the flux at each point on the image would be a sum of the flux scattered by each dust sheet.
Using this new model yields a much better fit (Figure~\ref{fig:rslt2_2} right column), with a dust sheet width \(\sigma_{\rm dust}\) of 12.4 days (for both sheets), as can be seen in Fig.~\ref{fig:rslt2_2}.

\begin{figure*}
\epsscale{0.9}
\includegraphics[width=\textwidth]{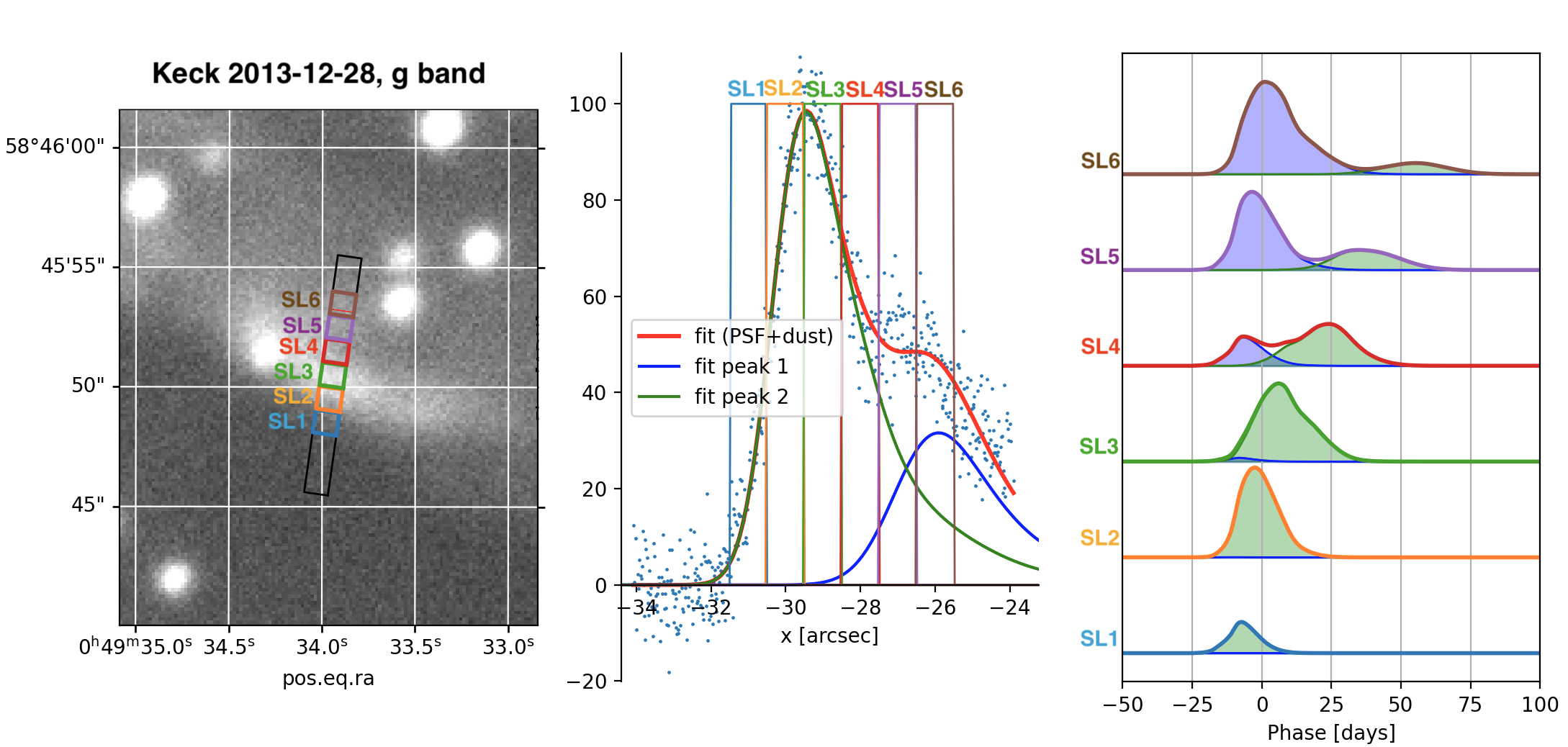}
\figcaption{Case 2: Double peak - effective light curves for different extraction areas along the slit. \textit{Left panel}: image of the LE, black box marks the profile extraction area, and the six colored boxes SL1 through SL6 mark prospective extraction areas for spectroscopy. \textit{Middle panel}: LE profile data and the LE profile model fit in red, with the two constituent LE profile models. {\it Right panel}: total effective light curves with contributions from both sheets. Line color corresponding to extraction area color on the left panel.  Blue and green shaded areas represent contribution from first and second peak, respectively.\label{fig:rslt2_double_peak_multisltlt}}
\end{figure*}

An interesting result in this case is how the effective light curve depends on the choice of extraction area along the slit. Repeating the same process for different extraction areas, we see how the relative contributions of the two overlapping dust sheets to the effective light curve transition continuously from the first sheet to the second, as shown in Figure~\ref{fig:rslt2_double_peak_multisltlt}.

\subsection{Case 3: Composite Slit}
\label{sec:rslt_3}

\begin{figure*}
\epsscale{0.9}
\includegraphics[width=\textwidth]{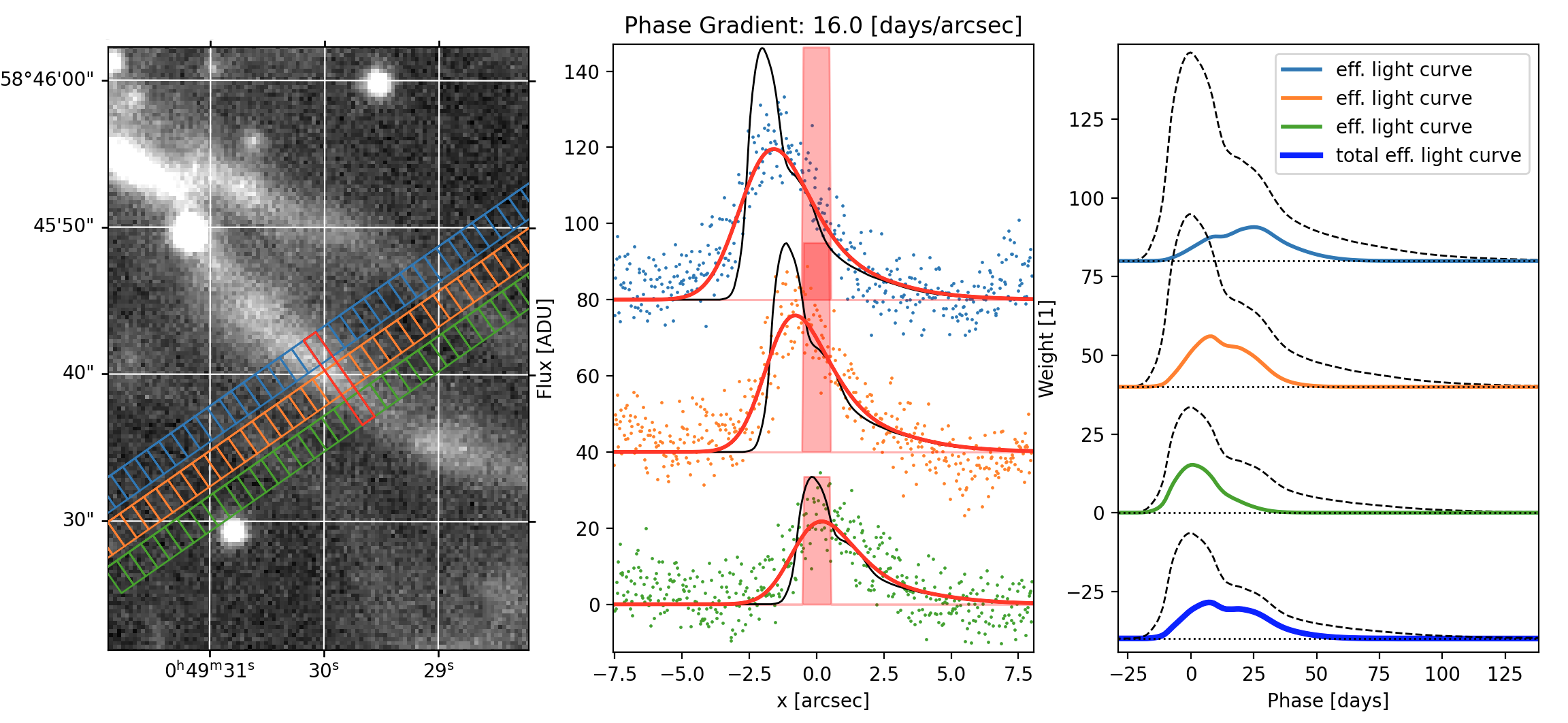}
\figcaption{Case 3: Composite slit modeling. \textbf{Left panel}: image used for modeling, the slit (thick red box) and three profile boxes marked by blue, orange and green ladder-shaped markings. The slit is almost parallel to the LE crest. It is divided into three slitlets (intersection of the slit with each of three the profile boxes). \textit{Middle panel}: LE profiles for the three profile boxes, with profile model fits (thick red lines) and the corresponding spectra extraction areas in the slit marked by a light red shaded box. \textit{Right panel}: effective light curves for the three extraction areas (light blue, orange, green), and the total effective light curve for the entire slit (thick blue).
Note, For example, how the north-eastern slitlet (light blue plot) contains contribution mainly from later phases, while the south-western slitlet (green plot) contains mainly phases near the peak.
\label{fig:rslt3_1}}
\end{figure*}


In LE spectroscopy, when measuring a faint but extended LE, it is a common practice to orient the slit nearly parallel to the LE crest and sum the spectra measured along the slit. This method increases the signal-to-noise of the measured spectrum, while introducing some complications to the process of spectral modeling.
The first complication comes from the increased inhomogeneity of the dust enclosed by the nearly parallel box. This means the light entering the slit comes from relatively distant and parts of the LE that may have significantly different properties (e.g., dust density, sheet width, inclination). 
Secondly, unless the slit is exactly parallel to the LE crest, each part of the slit will probe a different range of epochs of the light curve (see example in right panel of Figure~\ref{fig:rslt3_1}). We resolve these complications by first generating independent window functions and effective light curves for each part of the slit, then assigning a normalization factor for each part, and finally summing them together to produce a single effective light curve for the entire slit.

As an example, we take again the four image series from Case 1 as our input data, only now the profile box is set to be nearly parallel to the LE crest (Figure~\ref{fig:rslt3_1} left panel, red box enclosing blue dots). Apparent motion estimation was performed for the six new perpendicular profile boxes (colored boxes), and the result was \about16 days~arcsec$^{-1}$ for all six. We then focus on analyzing the LE in the third image (Keck LRIS, 2012-09-18), this time with a 1~arcsec wide near-parallel slit (Figure~\ref{fig:rslt3_1}, left panel). The total 6~arcsec long extraction area on the slit is shown by the red box, lying almost parallel to the LE crest. The total measured spectrum is a sum of six portions, one from each part of the extraction area (we divide the total extraction area into six slitlets, each 1\arcsec\ long).
For each of the six slitlets, we repeat the process from in Case 1: perform a free fit for each of the four images in the series to find peak locations which give us apparent motion, then perform a constrained fit using the corresponding projection scale to get the other parameters (e.g., dust width and peak flux) and calculate the effective light curve.

Now, we examine  our chosen image, Keck/LRIS 2012 September 18. We have an effective light curve for each of the six slitlets comprising the total extraction area (Figure~\ref{fig:rslt3_1}, right panel). 
We want to forward-model the spectrum that was measured by summing over the entire extraction area (red box), which contains six slitlets with six different effective light curves. But, the effective light curves are all self-normalized: the effective light curves of the first and sixth slitlets, for example, would look exactly the same even if the dust density in slitlet 1 would be a thousand times greater than the density on slitlet 6. In this hypothetical case, simply summing the effective light curves as they are would clearly produce a wrong model spectrum, since we would have same order of magnitude contributions from slitlet 6 (phases around 0 days) and from slitlet 1 (phases around 12 days), although the dust density difference would cause slitlet 1 to dominate the total measured spectrum.

To account for this effect, we multiply each effective light curve by a normalization factor before summing them all up to create the total effective light curve. The normalization factor is the peak flux obtained by the constrained fit, i.e., the height of the clean light curves (Figure~\ref{fig:rslt3_1} middle panel, black solid lines).


%% file: 01Chapters/05Discussion.tex
\section{Discussion}
\label{sec:disc}

		

Light echoes are powerful tools to obtain data of ancient transients with modern instrumentation. They pose a unique opportunity to characterize historical events in detail and compare them directly to the large spectrophotometric data sets of contemporary transients. However, if we want to go beyond simple spectroscopic classification, it is imperative to remove geometric and observational effects from the observed LE spectra. In the previous sections, we have shown how this can be done with forward modeling. In this section, we will discuss assumptions, requirements, limitations, and caveats of our method, as well as future work to apply it to several science objectives.

\subsection{Review of the method}
\textbf{The unique value of our method.} The transient's age and distance as well as the scattering dust filaments geometry (in particular the distance and orientation) can have direct impacts on the light echo that are often degenerate. The problem is that each of these parameters are often poorly known, if at all. The strength of our approach is that it is not necessary to determine them separately and independently. 
Our method assumes that these effects 
combine such that the light profile along a chosen direction is simply the projected light curve of the transient smeared by a Gaussian kernel. The only things that needs to be determined empirically are the projection scale and the width ($\sigma$) of the Gaussian kernel, which can be done with a time series of images. This is the single most important strength of our approach: without any knowledge about the transient's age, distance, or the dust filaments geometry and using only time-series imaging, we can predict which light curve phases contribute to an observed light echo spectrum.
Still, this method is not error proof, and one must take additional independent and measurable sanity checks into account when using it. We elaborate on important points and assumptions to bear in mind in the following.

\textbf{Apparent motion consistency.} 
A major assumption of our method is that the scattering dust filament can be well approximated by a planar sheet on length scales that are comparable to the time scale of the transient.
This requirement ensures that there is a simple linear conversion between spatial positions on the LE profile to phases on the light curve of the source event.
The simplest case is of course if the apparent motion is constant over the time period of the time series imaging so that it can be fit with a straight line. 
However, this can even work if the apparent motion is not constant, since our goal is not to determine the apparent motion over the full time span of the imaging, but to determine the apparent motion---and hence the corresponding projection scale---at the time of a given spectroscopic observation. For example, 
the apparent motion in a several-years or even decade-long time series of images may be adequately modelled 
with a quadratic (or even a higher-order) polynomial as long as it 
doesn't change dramatically over the time scale of the transient (typically months, not years). 
Fortunately, with a sufficient number of images in the time series, 
it is obvious when this assumption breaks down.

Even if the apparent motion condition is fulfilled, the scattering dust filament must also have a constant density and width over the spatial scales that are comparable to the light-travel distance for the timescale of the transient.
Otherwise, different phases will be scattered with unpredictable relative intensities, making it infeasible to model the light echo profile as a projected light curve. 
Visually, this assumption may seem to be violated often, considering the complicated structure apparent in many light echoes (e.g., in Fig~\ref{fig:LESpc_primer} one can see spacial variations in dust density over the extent of the slit). However, this is somewhat misleading: while there are significant surface brightness variations across the light echoes, it is often remarkably consistent over short time scales such as weeks or months that are similar to the those of most transients. Nevertheless, there are cases where these assumptions do break down, and we cannot forward model the light echo spectra. We therefore recommend caution when defining and planning spectroscopic observations, for example, to not simply target the brightest portions of light echoes, but those that also show similar structure in previous imaging. 

\textbf{Profile box orientation.} Another important decision that needs to be made when planning spectroscopic observations is the orientation of the slit with respect to the light echo crest (see  Fig.~\ref{fig:LESpc_primer}). Intuitively, the 
majority of observers would choose a slit that is set along the crest of the LE, which maximizes the flux captured by the slit. This is the best choice when the LE is very faint and it is the only way to obtain a usable spectrum. For bright LEs though, a slit roughly perpendicular to the light echo crest can be advantageous for the following reasons: 
\begin{itemize}
    \item If {\bf all} LE flux is summed up for the spectrum, then the spectrum is simply the lightcurve-weighted sum of all phases as long the dust geometry is not overly complicated. No forward-modeling is needed, in contrast to the spectra from a slit parallel to the light echo crest. This is the simplest possible analysis case. 
    \item Under favorable conditions (sufficiently thin scattering dust filaments and good observing conditions), a spectroscopic time series can be obtained if spectra of small slitlets are extracted independently (see Section~\ref{sec:rslt_2}).
\end{itemize}
This choice of slit position therefore strongly depends both on the scientific goals, the observing conditions, and the characteristics of the LE.

\textbf{Phase spread by dust and optics.} One of the key aspects of the analysis is to determine how much the template light curve needs to be smeared out to fit the light echo profile. This smearing is the combined effect of the dust width, $\sigma_{\mathrm{dust}}$, and from observational effects like seeing, $\sigma_{\mathrm{PSF}}$. Ideally, the amount of smearing is fitted using high S/N imaging taken directly before or after the spectroscopic observations, which allows for its determination presumably under the same observing conditions. The assumption whether the observing conditions (mainly seeing) did not change can be tested if there is imaging both before and after the spectrum is taken. This is the most direct and robust way to determine the smearing out. 

However, for various reasons, it is not always possible to obtain imaging before and/or after the spectroscopic observations. In that case, the smearing out factor needs to be determined in a more indirect way: first, $\sigma_{\mathrm{dust}}$ 
needs to be determined by fitting a high S/N light echo profile at any epoch (see Section~\ref{sec:recipe_step3} for an explanation of the fitting procedure). 
Then, with the assumption that this dust width does not change between the epochs and an estimate of the seeing during the spectroscopic observations obtained from another source (e.g., seeing monitor or photometry of low S/N imaging), we can estimate the total smearing, $\sigma_{\mathrm{tot}}$, at the time of the spectroscopic observations. This is of course a more involved procedure that relies on several assumptions, and is therefore less desirable. 

\textbf{Dust density variations.} Most LEs come in groups of arclets that share a common apparent motion. Over time spans of years, however, these LE groups do undergo changes (fading, brightening, morphology changes) in an apparently messy way. On a closer look, these LE groups seem to come from individual filaments that are aligned in parallel planes. Sometimes these filaments are so close that the resulting LEs overlap on the sky, and we emphasize the importance of always verifying that fits to LE profiles are reasonable. For example, in Case 2 (Section~\ref{sec:rslt_2}), we were unable to obtain a good fit of the LE profile using the projection scale derived from the apparent motion. It took several iterations and checking for consistency in different filters until we found that this is the combination of two LEs from distinct but closely associated dust filaments. It is clear that not all LEs can be forward modeled in straightforward way. In these cases, the LE spectra cannot be used for a detailed analysis and characterization, but we find that this is typically a small minority.


\subsection{Next steps}
Here, we provide a brief review of several prospective research objectives where our method is particularly valuable. 

\textbf{Explosion asymmetry.} Nearby SN remnants for which we have observed light echoes can benefit from the ability of this method to discriminate between LE spectrum features that arise from dust and instrumentation characteristics, and those that arise from intrinsic variation in the SN explosion from different observing directions. This will allow researchers to relate distinct observable features in the remnant directly to differences in spectral features of the explosion as seen from different lines of sight. 

\textbf{Temporally Resolved LE Spectra}.
One of the most exciting opportunities LEs present is temporal resolution of 3D spectroscopy (i.e., spectra attained for lines of sight different than the Earth's) for historical and contemporary SNe.
Temporal resolution is achieved by orienting the slit perpendicular to an LE crest, so that different parts of the slit probe different epochs of the SN light curve.
The difficulty in aligning the slit this way is that it significantly reduces the amount of light within the slit. However, it is possible to increase the signal-to-noise by using a wide slit, potentially at the expense of wavelength resolution or span, depending on the instrumentation. The spatial-variance-related issues involved in wide slit LE spectroscopy (dust density and phase changes across the slit width) can be overcome by spatially resolved modeling, as demonstrated in Section~\ref{sec:rslt_3}.

\textbf{Subtyping.} Classifying historic SNe into types has been previously achieved with light echoes. Detailed subtyping, however, has not been done thus far as it requires a significantly greater level of precision. Using our method, this level is achievable thanks to the ability to temporally resolve LE spectra to compare to growing spectral libraries of contemporary SNe.

%% file: 01Chapters/06Conclusion.tex
\section{Conclusions}
\label{sec:conc}

LEs offer the only opportunity to observe ancient transients with modern instrumentation. They have been used to spectroscopically classify Galactic SNe like Tycho's SN, and view the same transient from multiple different directions (e.g., Cas\,A and SN\,1987A). However, in order to take the next step and identify more subtle spectral features, it is necessary to separate observational effects from astrophysical signatures. 

In this paper, we have presented a method for proper forward modeling of LE spectra, which utilizes three key concepts: (1) the LE profile is the projected light curve of the source transient, (2) the LE apparent motion can be used to derive the LE projection scale, and (3) properties of the scattering dust filament like the dust width smear out the projected light curve. With this method, a synthetic LE spectrum is constructed that accounts for the geometry of the scattering dust filament as well as observational effects like seeing and slit width/orientation. This makes it possible to directly compare spectral templates derived from a spectrophotometric transient library to the observed LE spectra. No assumptions about age and distance of the source transient are necessary, which is crucial since these are often not known to high accuracy. We derive how slit orientation and width can be modeled and accounted for, and show how the method can be applied to three different examples of Tycho's LEs, each with its own challenges and peculiarities. These cases illustrate how to go beyond the simplest application of the technique to account for scenarios involving complicated dust structures (e.g., overlapping dust sheets) or a wide slit. When combined with well-chosen observing strategies, our method will allow for detailed and accurate LE modelling for the vast majority of cases across a broad range of astrophysical transients.

\section*{Acknowledgement}

We are very thankful for the support of R.P., R.R.-H., J.J., and X.L. by NSF AST grants 1814993 and 2108841.
R.P. and  D.P. acknowledge support from Israel Science Foundation (ISF) grant 541/17.
The UCSC team is supported in part by NASA grant NNG-17PX03C, National Science Foundation (NSF) grant AST-1720756, the Gordon and Betty Moore Foundation, the Heising-Simons Foundation, and fellowship from the Alfred P. Sloan Foundation and the David and Lucile Packard Foundation to R.J.F.
C.D.K. is partly supported by a CIERA postdoctoral fellowship.
A.V.F. was supported by the Christopher R. Redlich Fund and numerous donors.
The work of S. Margheim is supported by NOIRLab, which is managed by the Association of Universities for Research in Astronomy (AURA) under a cooperative agreement with the National Science Foundation.
M.R.S. is supported by the STScI Postdoctoral Fellowship.


Some of the data presented herein were obtained at the W. M. Keck Observatory, which is operated as a scientific partnership among the California Institute of Technology, the University of California and the National Aeronautics and Space Administration. The Observatory was made possible by the generous financial support of the W. M. Keck Foundation. The authors wish to recognize and acknowledge the very significant cultural role and reverence that the summit of Maunakea has always had within the indigenous Hawaiian community.  We are most fortunate to have the opportunity to conduct observations from this mountain. 

\section*{Software and third party data repository citations} \label{sec:cite}

\facilities{Mayall (MOSAIC-1 wide-field camera), Mayall (Mosaic-3 wide-field camera), Keck:I (LRIS), Keck:II (Deimos)}


\software{
\texttt{astropy} \citep{astropy:2013,astropy:2018,astropy:2022},
\texttt{matplotlib} \citep{Hunter:2007}, 
\texttt{SciPy} \citep[][]{2020SciPy-NMeth}, 
\texttt{NumPy} \citep{harris2020array},
\texttt{LMFIT} \citep{matt_newville_2023_8145703}, 
\texttt{pandas} \citep{reback2020pandas, mckinney-proc-scipy-2010},
\texttt{Photpipe} \citep{Rest05a,Rest14},
\texttt{Dophot} \citep{Schechter93},
\texttt{hotpants} \citep{2015ascl.soft04004B} \footnote{https://github.com/acbecker/hotpants},
\texttt{Swarp} \citep{2002ASPC..281..228B},
}

%% file: 02Appendices/01CC1.tex
\section{Core Concept 1: LE Profiles are Projected Light Curves}
\label{sec:XCC1}

\begin{figure*}
\epsscale{1.2}
\includegraphics[width=\textwidth]{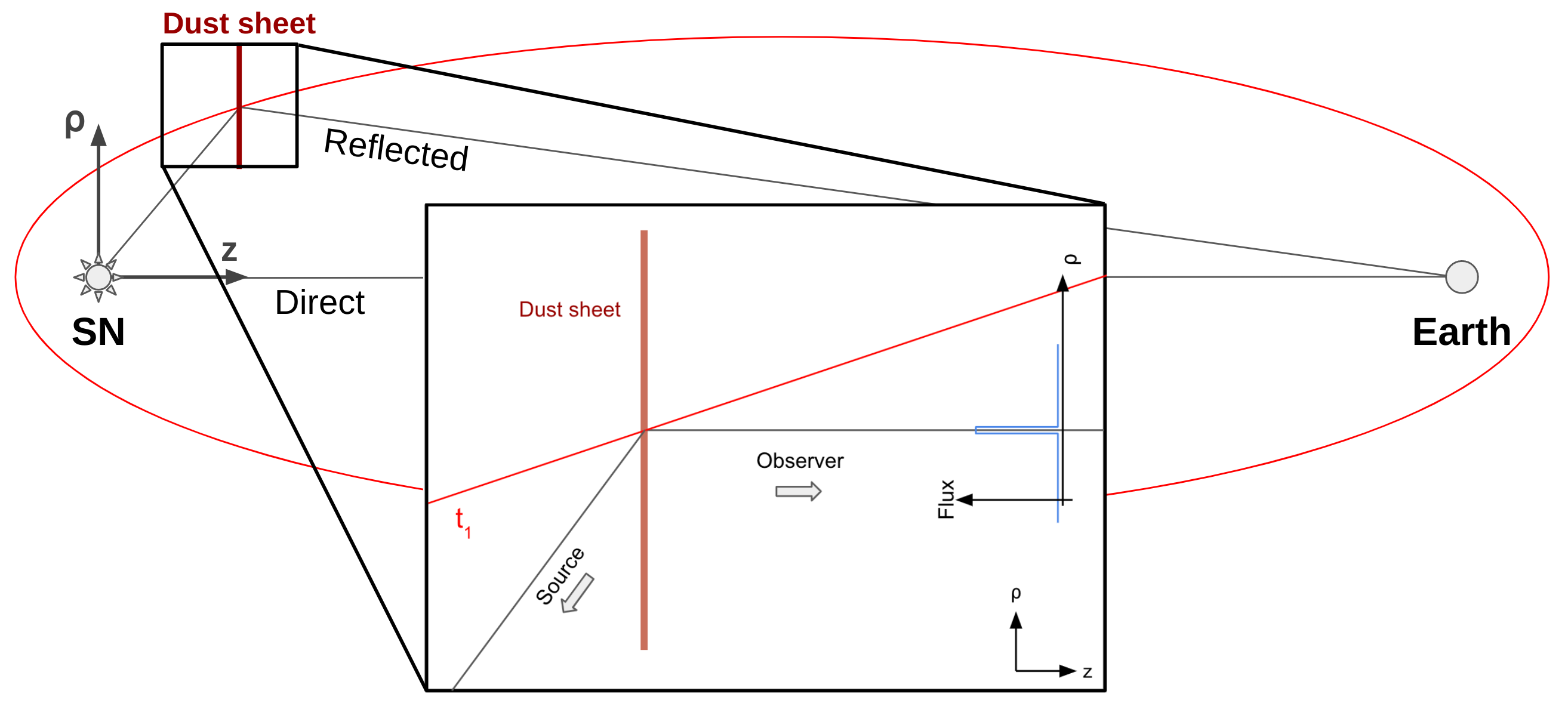}
\figcaption{Ellipsoid of potential scattering sites (cross section through line of sight), for a given time difference between observations of direct and scattered light. Dust located in any of these locations will cause an LE. A thin dust sheet is illustrated in light brown. The part of the dust sheet which intersects the ellipsoid is scatters light from the SN, which is observed as an LE on earth. \textit{Inset:} Zoom in on scattering area, along with the flux profile of the LE for an impulse light curve.
\label{fig:ellipsoid}}
\end{figure*}

This section is intended to provide reasoning in support of core concept 1, which was described in Section~\ref{sec:mthd_CC1} above.
We first assume the idealized case of an impulse transient source: all light is emitted over a very short time interval---essentially a Dirac delta function light curve. In such a case, the LE appears as a thin arclet (or, if the dust sheet is sufficiently large, as a thin ring surrounding the source event). This is illustrated in Figure~\ref{fig:ellipsoid}. 
As shown in the figure, the image of the LE taking the shape of an arclet will have an LE profile in the shape of a single impulse.

\begin{figure}
\epsscale{1.2}
\includegraphics[width=\linewidth]{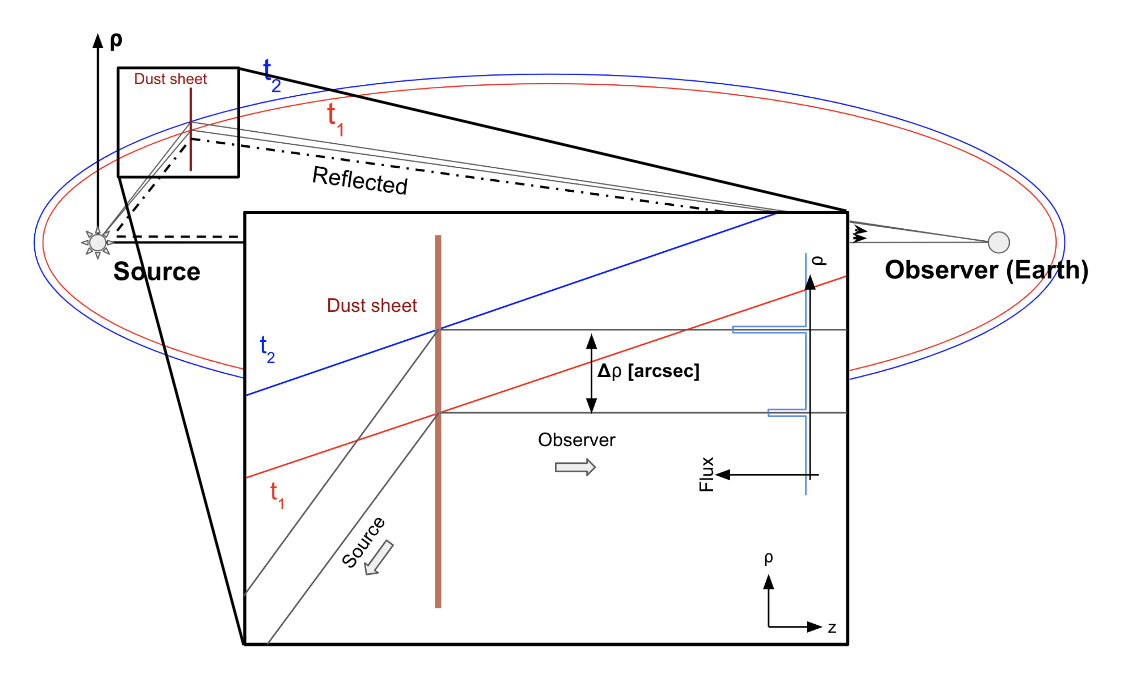}
\figcaption{Similar to right panel of Figure~\ref{fig:ellipsoid}, this time the light curve is composed of two impulses, with their corresponding time delays between arrival of direct light and LE.
Note that $t_2>t_1$, meaning that $t_2$ corresponds to the first peak (i.e., whose direct light was observed first) than that of $t_1$, and since both events are observed on the same date the time delay is greater for the earlier event. 
\label{fig:CC1_double_impulse}}
\end{figure}

We examine another idealized light curve, one with two impulses (Figure~\ref{fig:CC1_double_impulse}).
Viewed from earth, the LE appears as two concentric arclets. The outer arclet (greater $\rho$) corresponds to the first impulse and the inner arclet corresponds to the later impulse. Taking the flux profile that ``cuts through'' the LE, we obtain a double impulse profile.

\begin{figure*}
\epsscale{1.2}
\includegraphics[width=\textwidth]{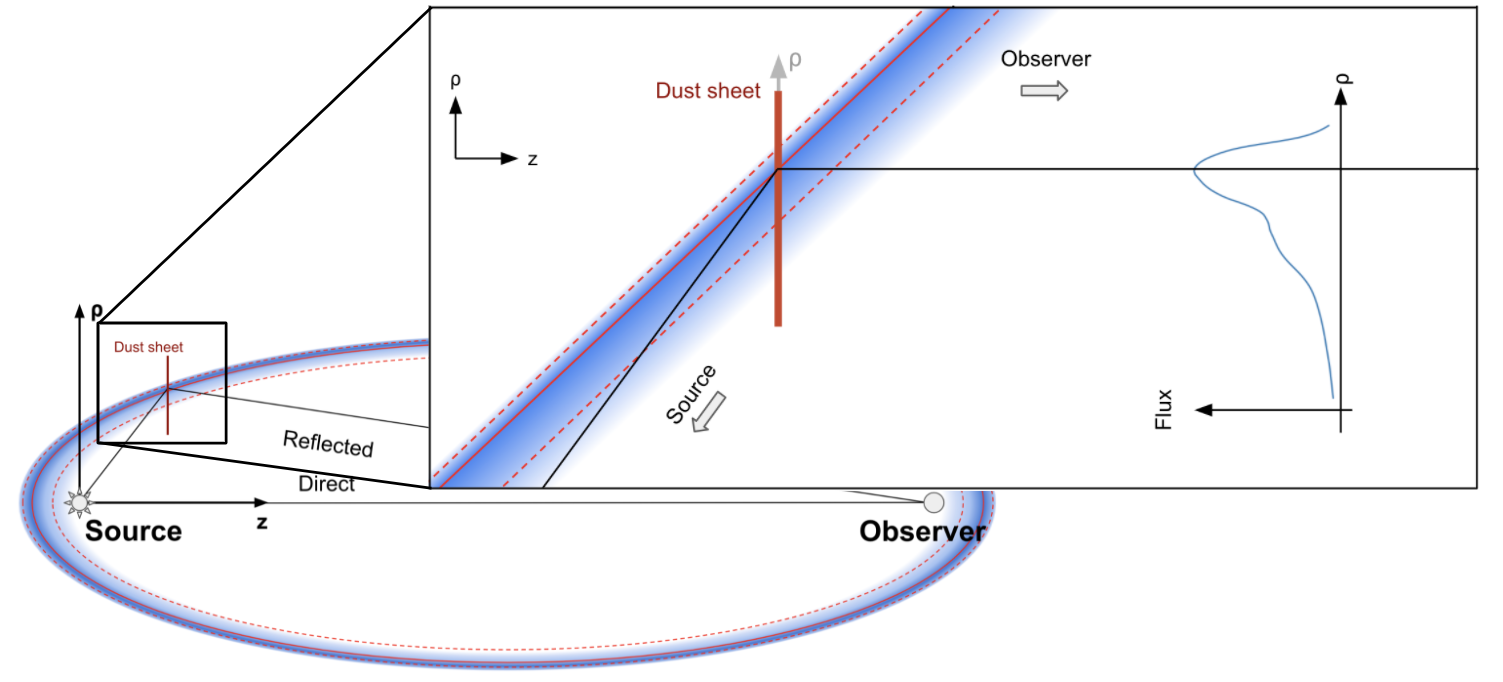}
\figcaption{An LE profile is a projected light curve. The blue-shaded region indicates the ellipsoidal shell containing a continuum of ellipsoids for a range of phases. The shade level represents the brightness of the light curve at the respective phase. This then determines the brightness of the light scattered from that location (assuming constant dust density, for simplicity. Note that we also neglect the inverse-squared distance dependence of light emitted by the source). The rise of the light curve, i.e., early phases, is located at larger separation $\rho$ than the fall. This is because the part of the dust sheet farthest away from the SN remnant (greater $\rho$) always scatters the earliest phases of the light curve, since those phases have had more time to propagate away from the event.\label{fig:Im_FluxProf}}
\end{figure*}

Lastly we move on to an actual light curve, that of a typical SN\,Ia, which has a time span on the order of 100 days. Its light curve can be thought of as a continuous set of delta functions along the phase axis, normalized by the intensity of the light curve at each phase. We consider the continuous set of LE ellipsoids associated with each of these delta functions (an ellipsoidal shell, Figure~\ref{fig:Im_FluxProf}). 
For a given observation date, each of the ellipsoids has a slightly different time delay $t$, corresponding to a specific phase in the light curve, and hence a different size. Assuming an infinitely thin dust sheet intersecting all ellipsoids in the shell, we have that the outer ellipsoids (greater $t$) intersect the dust sheet at a greater angular separation $\rho$ from the $z$ axis. This in turn means that each phase is projected onto a slightly different position on the sky. The total LE can then be described as a continuous set of arclets, or alternatively as a thick arclet whose surface brightness is a function of angular separation $\rho$.
Examining the LE profile shown in Figure~\ref{fig:one_big_figure} (top left box, Core Concept I), we find that it is in fact a projection of the light curve,\footnote{N.B. that in Figure~\ref{fig:one_big_figure} (top left box - Core Concept I) we have a real world example showing also the effects of a finite dust width and observing conditions, which is discussed in Section~\ref{sec:mthd_CC2} and Section~\ref{sec:mthd_CC3}.} in this case an SN Ia. The typical asymmetric shape of the peak is easily discernible.

We now define a new important parameter, the projection scale, designated $P$.
\begin{definition}
    The \textbf{projection scale} of a given LE profile is the ratio between the angular separation $\Delta \rho$ between projected locations and time difference $\Delta t$, of any two phases on the source light curve.
\end{definition}
In practice, the projection scale serves as the coefficient for converting from phase to angular separation in the projection of the light curve to the LE profile. For example, for the case show in Figure~\ref{fig:CC1_double_impulse} the projection scale is $P=\frac{t_2-t_1}{\Delta\rho}$, and is typically measured in units of days~arcsec$^{-1}$.

It is worth noting that the projection scale $P$ strongly depends on the angle of the dust filament. For example, consider a dust filament or sheet aligned orthogonal to the line of sight and contained in a plane going through the source SN.
In that particular case, $P$ is simply the projected light speed at the distance of the dust filament. For example, for a dust sheet located at $z=0$ distance of $D=10,000$~ly from the observer, $P$ would be $1\frac{\rm light-day}{c} / \frac{\rm light-day}{10,000~{\rm ly}} = \frac{1~{\rm day}}{0.056~{\rm arcsec}} = 17.7$~days~arcsec$^{-1}$. However,  if the scattering dust filament is inclined ``backward'' or ``forward'', $P$ will vary accordingly and the projected light curve is ``squished'' or ``stretched'' compared to our nominal orthogonal scattering dust filament, respectively  (see Figure~\ref{fig:proj_LC}).

\begin{figure}
\epsscale{1.2}
\includegraphics[width=\linewidth]{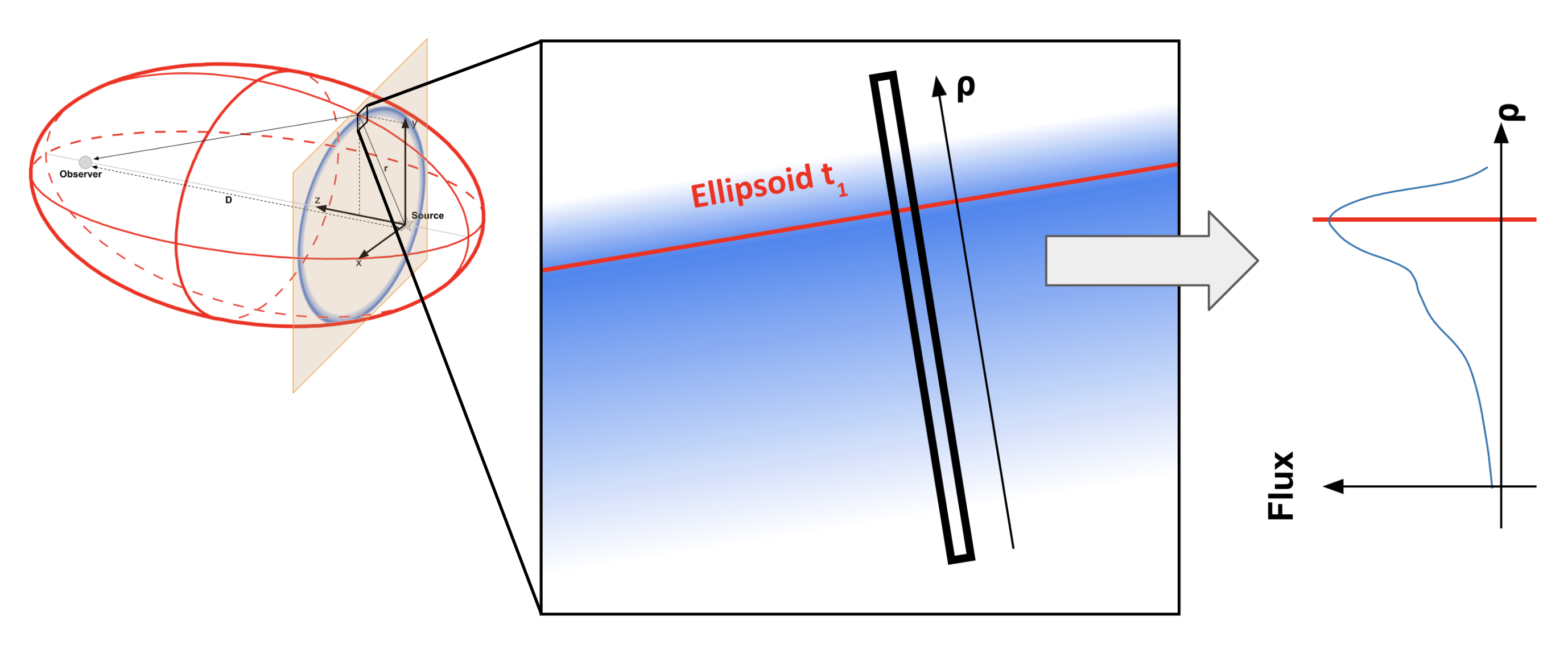}
\figcaption{Core Concept 1: an LE profile is a projected light curve. Suppose for the case illustrated in Fig.~\ref{fig:LE_geometry} that we observe an image whose FOV is only a small part of the entire LE ring (marked as black square in the dust sheet plane, on the left). \textit{Middle:} an inset zoom in of the observed image. The LE is illustrated in shaded blue. \textit{Right:} taking an LE profile from the image, we get a projected light curve.
\label{fig:XCC1}}
\end{figure}

%% file: 02Appendices/02CC2.tex
\section{Core Concept 2: The Projection Scale is Inverse to the Apparent Motion}
\label{sec:XCC2}

Under certain conditions, there is a simple and well defined relation between the two:
\begin{equation}
    P = \frac{1}{\mu} \label{eq:P_V}
\end{equation}
Simply put, the conditions for this relation to hold are that the dust filament is well approximated by a planar sheet with constant density for the length scale associated with either the transient duration or the time interval between the first and last observations, whichever is greater. For example, SNe have typical time scales of 100 days, and the relation between apparent motion and projection scale holds for observations taken within this time frame if the dust filament is approximately planar with constant density within 100 light-days.
Note that in this section we still assume a thin dust structure and no degradation by atmosphere or optics, the effects of which we defer to Section~\ref{sec:mthd_CC3}.

We consider an LE that is observed at three time delays $t_0$, $t_1$, and $t_2$ measured from the date when light from the peak first reached Earth (Figure~\ref{fig:appmotion_illu}).
In each observation, the crest (peak brightness) of the LE appears where the dust structure intersects the ellipsoid corresponding to the time delay of that observation. Since the ellipsoids' radii increase over time, we get the characteristic outward motion of the LE away from the source event.

Now consider phase +20 days after peak brightness, and observation at time $t_1$ (top middle panel). Light from this phase has had 20 days \textbf{less} time to propagate away from the SN compared to light from peak brightness phase, hence it's corresponding ellipsoid intersects the dust sheet closer to the SN. Since $t_1-t_0=20$~days, This intersection is \textbf{the same line} as the one where the peak brightness ellipsoid has intersected the dust sheet at the observation at time $t_0$.
Since the angular separation between these two intersection lines is 2 arcsec, we get that the projection scale is $\frac{20 days}{2 arcsec}=10 \frac{days}{arcsec}$, as phase +20 days is located 2 arcsec away from peak brightness on the LE profile. At the same time, the apparent motion of the LE is $\frac{2 arcsec}{20 days}=0.1 \frac{arcsec}{days}$, since in 20 days time the peak brightness crest has ``traveled'' 2 arcsec across the sky.
The equivalence of $\Delta t$, $\Delta \rho$ as being both intervals between two phases on a single LE observation, and at the same time intervals between the same phase on two different LE observations, is what establishes the relation in Eq.~\ref{eq:P_V}.
A real-world example can be found in Fig~\ref{fig:rslt1_1}, which is discussed in detail in Section~\ref{sec:rslt_1}. Such good linear fits over months and even years are common, and indicate that the scattering dust structures are indeed often well approximated by planar dust filaments.


Note that finding the location of the peaks can be achieved by fitting an LE profile model (as defined in Section~\ref{sec:recipe_step3}) with an unconstrained projection scale, that is a model where the projection scale mapping the template light curve to spatial coordinates is left as a free parameter. This is valid since the peak location is not affected much by the value of the projection scale and other parameters of the model (such as projection scale and dust width), and thus we can get a good estimate of the peak location even if all the other parameters are incorrect.

We have shown that the projection scale, the most important ingredient for the analysis of light echo spectroscopy, can be determined solely by measuring the apparent motion of the light echo. This works for arbitrary orientations of the light echo profile box, as long the apparent motion of the light echo within this box is consistent and constant. This is one of the main differences to previous work: instead of trying to determine or model the full scene of the transient (age, distance) and the scattering dust (distance, inclination), we empirically determine the apparent motion, which is a differential measurement of time and position that can be directly converted to projection scale. This is the only parameter needed to interpret and analyze the observed light echo. 


\begin{figure*}
\epsscale{1.2}
\includegraphics[width=\textwidth]{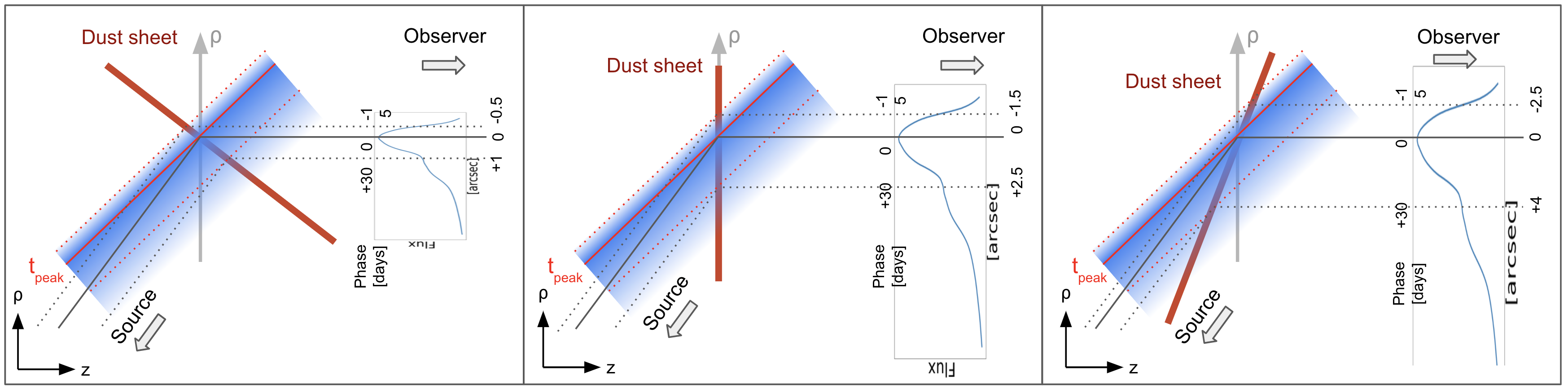}
\figcaption{The LE profile depends on the inclination of the scattering dust filament. From left to right, the dust filament is inclined backwards, orthogonal, and forward with respect to the $z$-axis (line of sight), respectively. The red solid line indicates the ellipsoid $t_{peak}$ associated with the transient at peak, whereas the red dotted lines indicate the ellipsoids associated with phases -15 and 30 days. Note how the change in inclination angle changes the location of the intersection of the -15/+30 ellipsoids with the dust sheet on the $\rho$ axis, thus increasing the angular separation between those phases on the projected light curve. The projected light curve is thus stretched as the dust filament inclination changes from backward to forward inclination. \label{fig:proj_LC}}
\end{figure*}

%% file: 02Appendices/03CC3.tex
\section{Core Concept 3: The Smearing effect of Dust Width and Optical PSF Modeled as Convolution}
\label{sec:XCC3}

\textbf{Dust:} 
The effect of a thick dust width is shown qualitatively in Figure~\ref{fig:dust_width}. Simply put, when the dust is no longer thin, any given ellipsoid probes the filament not in a single location, but rather over a range of locations that are thus projected onto a range of locations on the sky plane of the observer. This introduces a smearing effect to the LE profile, that is not present in the light curve.
We assume that the dust distribution is well approximated by a Gaussian function of width $\sigma_{0,{\rm dust}}$. Hence, the smearing effect is modeled as a convolution with a Gaussian function $G(\sigma_{\rm dust})$ of width $\sigma_{\rm dust}$
\\\\
\textbf{PSF:} 
In addition to the effects of dust width, the LE profile is convolved once more with a point-spread-function (PSF) due to the effects of telescope optics and atmospheric conditions at the time of the observations. The effect of the PSF on the LE profile is modeled by a another Gaussian function, marked $G(\sigma_{\rm PSF,eff})$.\footnote{$\sigma_{\rm PSF,eff}$ is the effective PSF. It's relation to the actual measured PSF is a cosine factor (Eq.~\ref{eq:PSF_eff}), as explained in Appendix~\ref{sec:XCC3}.}
\\\\
The combined effect of dust and PSF is the modeled by the convolution of their separate kernels. Since both kernels are Gaussian functions, the result of taking their convolution with each other is also a Gaussian, marked $G(\sigma_{\rm tot})$. The width $\sigma_{\rm tot}$ of this combined effect Gaussian is given by the hypotenuse sum of the separate widths (Eq.~\ref{eq:sig_tot}). Ultimately, the relation between the observed LE profile $F_{\rm LE}$ and the event's light curve $F_{\rm LC}$ is given by the following equations:
%
\begin{eqnarray}
F_{\rm LE} & = & F_{\rm LC} \ast G(\sigma_{\rm tot}) \label{eq:LE_proj_prof}\\
G(\sigma_{\rm tot}) & = & G(\sigma_{\rm dust}) \ast G(\sigma_{\rm PSF,eff}) \label{eq:phs_smear} \\
\sigma_{\rm tot}^2 & = & \sigma_{\rm dust}^2 + \sigma_{\rm PSF,eff}^2 \label{eq:sig_tot}
\end{eqnarray}
It is important to note that the dust width effect kernel $G(\sigma_{\rm dust})$ is not only dependent on the actual geometric width $\sigma_{0,{\rm dust}}$ of the dust filament, but is also modified by the angle of the dust filament and the LE ellipsoid with respect to the line of sight, which in turn depend on the age of the event and the distance from the observer. In the vast majority of cases, these parameters are only known to a rough approximation, and it is therefore virtually impossible to determine $\sigma_{\rm dust}$ accurately enough a priori. However, since we have already obtained the projection scale, we need not bother with these sorts of estimations. This is because the projection scale allows us to switch directly between the spatial effect of the dust on the observed LE profile (characterized by $\sigma_{\rm dust} [arcsec]$) to the desired temporal effect (spectral phase blending) of the dust on the observed LE spectrum (characterized by $\sigma_{\rm dust} [days]$).
All that is left for us to do is to obtain an estimate of the spatial dust effect on the LE profile, i.e., $\sigma_{\rm dust} [arcsec]$. In the next section we show how we obtain this estimate directly from the observed LE profile, given that the light curve shape of the transient is known or at least constrained beforehand (which is usually the case).
\\\\    


We show an example of this in Figure~\ref{fig:rslt1_2} for an LE of Tycho's SN, which was classified as a normal type Ia SN \citep{Krause2008_TYC}. 
We use the measured apparent motion to determine the projection scale, with which we can seamlessly convert from arcsec to days and vice versa. From previous studies, we have an estimate that Tycho's SN was a normal SN Ia with $\Delta m_{15}=1.00$. 
\footnote{$\Delta m_{15}$ is the decline rate parameter for type Ia SNe, defined as $\Delta m_{15}(B)_{obs}=B_{obs}(+15 days)-B_{obs}(max)$ \citep{1993ApJ...413L.105P}.}

This means we can take a light curve from the SN Ia template library \citep{1996ApJ...473...88R} that matches the stretch of Tycho's SN and compare it against the light echo profile, using the projection scale for converting phase to arcsec
(see black line in the top right panel of Figure~\ref{fig:rslt1_2}).
We convolve the projected light curve with a Gaussian kernel $G(\sigma_{\rm tot})$ of variable width $\sigma_{\rm tot}$ until we find the width which best fits the observed light echo profile
(red line in top right panel of Fig~\ref{fig:rslt1_2}).
The only free parameters are $\sigma_{\rm dust}$, the position of the peak of the light curve and a flux scaling factor.
\\\\



\begin{figure}
\epsscale{1.2}
\includegraphics[width=\linewidth]{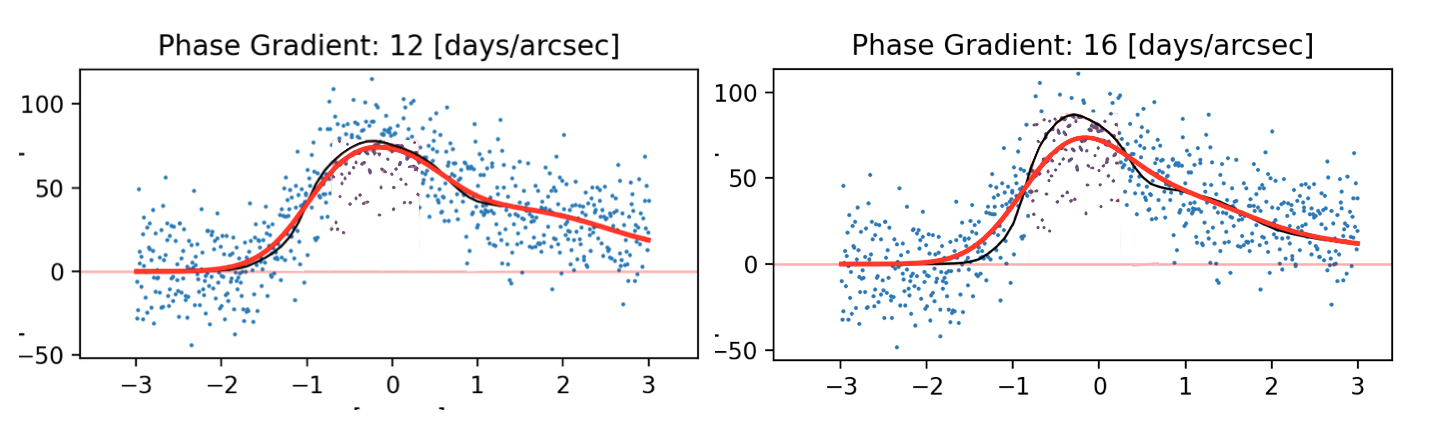}
\figcaption{Projection scale/dust width degeneracy. Both panels show the same LE profile data with a good fit ($1<\chi^{2}_{\nu}<1.1$), but with different parameter values. \textit{Left:} With an unconstrained projection scale (free fit), we obtain a good fit that is, however, inconsistent with the measured apparent motion. This fit also gives an unrealistically short dust width (about 1~day FWHM, effectively zero relative to the level of phase smear by the PSF). \textit{Right:} Using a phase gradient set by apparent motion also gives a good fit with a more plausible dust width (11~days FWHM).
\label{fig:phs_grd_dst_wid_deg}}
\end{figure}

\noindent {\bf Projection scale/dust width degeneracy:}
Our analysis method relies heavily on the estimation of projection scale from the measurement of apparent motion, using a series of images taken on consecutive epochs. The main reason for obtaining the projection scale in that manner rather than by curve fitting, is that a thick dust sheet can have a similar effect on the light echo profile to that of a small projection scale. This means that these two parameters are degenerate in a sense. This is demonstrated in Figure~\ref{fig:phs_grd_dst_wid_deg}, where the same light echo profile is fitted with two different models. Note that both models are very good fits, although they have significantly different projection scales.
\\\\

\noindent {\bf Effective PSF:} In general, the PSF of an observation is measured in terms of the Full-Width-Half-Maximum (FWHM). For our purposes, the PSF usually can be sufficiently characterized with a circular Gaussian having a single width $\sigma_{\rm PSF}$, as measured by standard PSF photometry.
We denote the angle between the the light echo profile box and the line perpendicular to the ``crest'' of the light echo as $\alpha$, such that $\alpha=0$ when the profile box is perpendicular to the crest. This is also the angle of maximum projection projection scale. 
Then $\sigma_{\rm PSF,eff}$ is related to $\sigma_{\rm PSF}$ with the following relation:

\begin{eqnarray}
\sigma_{\rm PSF,eff}  =  \frac{\sigma_{\rm PSF}}{\cos{\alpha}} \label{eq:PSF_eff}
\end{eqnarray}
 
\begin{figure}
\epsscale{1.0}
\includegraphics[width=\linewidth]{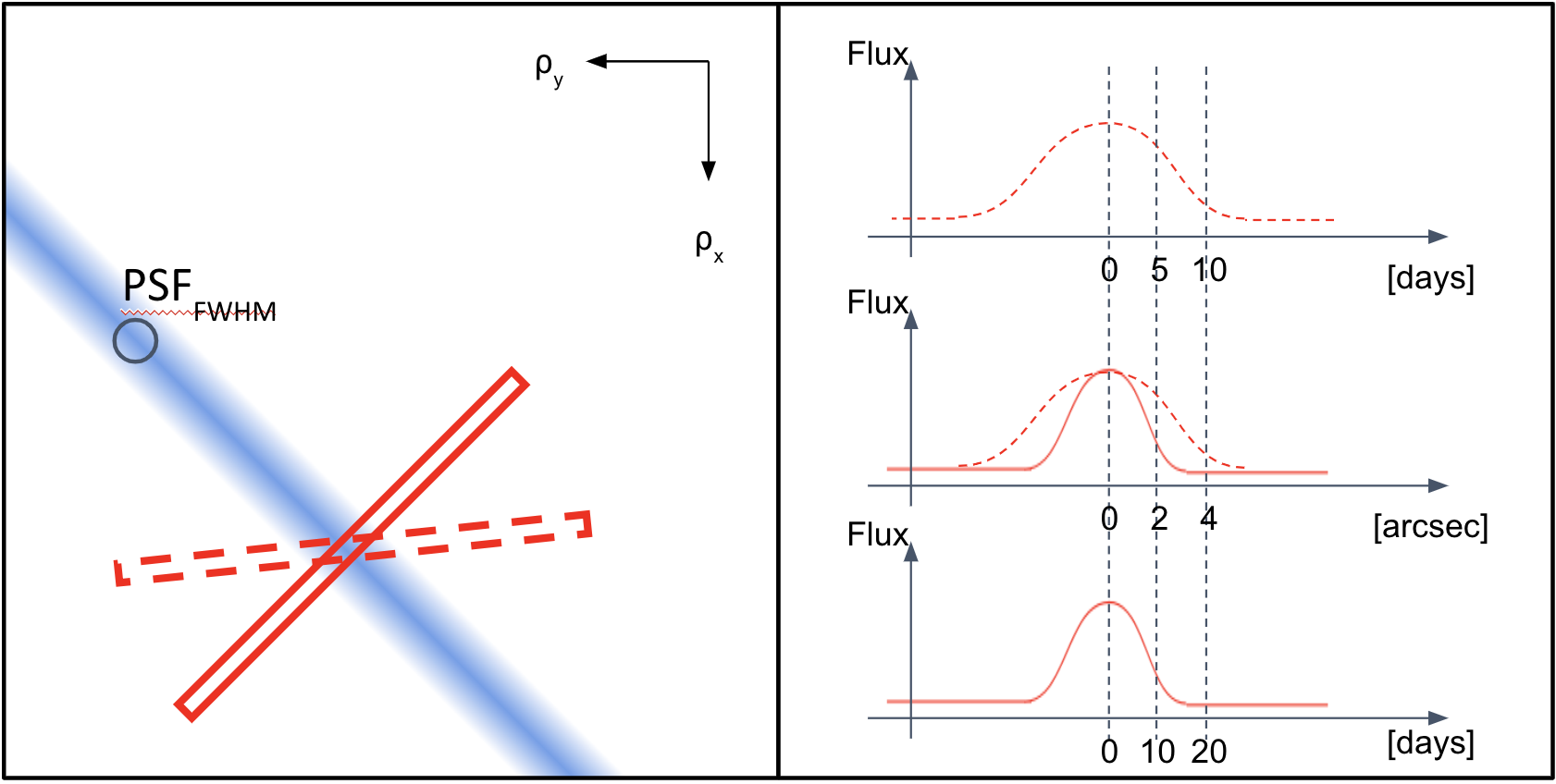}
\figcaption{Effect of position angle on effective PSF. For a given PSF (in this case a circular one for simplicity), it's effective width on the light echo profile will be wider for shallower angles with respect to the light echo crest. \label{fig:eff_psf}}
\end{figure}

We note that for small angles $\alpha$ between to LE profile box and the LE motion vector, we have $\sigma_{\rm PSF,eff} \approx \sigma_{\rm PSF}$. This is not a big effect, and needs to be taken into account only for $\alpha\gtrsim15\deg$ (see Figure~\ref{fig:eff_psf}).
This means that for cases with large $\alpha$, we must estimate the angle between the profile box and the motion vector in order to get $\sigma_{\rm PSF,eff}$ (see step 2 in Section~\ref{sec:method}).
